\documentclass[10pt,journal,compsoc]{IEEEtran}
\newif\ifpeerreview

\peerreviewfalse

\usepackage[nocompress]{cite}
\usepackage{url}
\usepackage{amsmath,amssymb,graphicx}

\usepackage{algorithmic}
\usepackage{graphicx}
\usepackage{textcomp}
\usepackage{xcolor}
\usepackage{array}
\usepackage{multirow}
\usepackage{float}
\usepackage{hyperref}

\usepackage[numbers,sort&compress]{natbib}

\usepackage{pifont}
\newcommand{\cmark}{\ding{51}}%
\newcommand{\xmark}{\ding{55}}%

\usepackage{xcolor}
\def\BibTeX{{\rm B\kern-.05em{\sc i\kern-.025em b}\kern-.08em
    T\kern-.1667em\lower.7ex\hbox{E}\kern-.125emX}}

\newcommand{\specialcell}[2][c]{%
  \begin{tabular}[#1]{@{}c@{}}#2\end{tabular}}

\usepackage{lipsum} 

\usepackage[switch]{lineno}

\newcommand{\paperID}{0016}

\title{Denoising Diffusion Probabilistic Model for Retinal Image Generation and Segmentation}

\author{Alnur~Alimanov  
        and~Md Baharul~Islam,~\IEEEmembership{Senior Member,~IEEE}
\IEEEcompsocitemizethanks{\IEEEcompsocthanksitem A. Alimanov is with the Department of Computer Engineering, Bahcesehir University, Istanbul,
Turkey, 34349.\protect\\
E-mail: alnur.alimanov@bahcesehir.edu.tr
\IEEEcompsocthanksitem Md B. Islam is with College of Data Science and Engineering, American University of Malta, Bormla, BML 1013, Malta.}
}

\begin{document}

\IEEEtitleabstractindextext{%
\begin{abstract}
Experts use retinal images and vessel trees to detect and diagnose various eye, blood circulation, and brain-related diseases. However, manual segmentation of retinal images is a time-consuming process that requires high expertise and is difficult due to privacy issues. Many methods have been proposed to segment images, but the need for large retinal image datasets limits the performance of these methods. Several methods synthesize deep learning models based on Generative Adversarial Networks (GAN) to generate limited sample varieties. This paper proposes a novel Denoising Diffusion Probabilistic Model (DDPM) that outperformed GANs in image synthesis. We developed a Retinal Trees (ReTree) dataset consisting of retinal images, corresponding vessel trees, and a segmentation network based on DDPM trained with images from the ReTree dataset. In the first stage, we develop a two-stage DDPM that generates vessel trees from random numbers belonging to a standard normal distribution. Later, the model is guided to generate fundus images from given vessel trees and random distribution. The proposed dataset has been evaluated quantitatively and qualitatively. Quantitative evaluation metrics include Frechet Inception Distance (FID) score, Jaccard similarity coefficient, Cohen's kappa, Matthew's Correlation Coefficient (MCC), precision, recall, F1-score, and accuracy. We trained the vessel segmentation model with synthetic data to validate our dataset's efficiency and tested it on authentic data. Our developed dataset and source code is available at \url{https://github.com/AAleka/retree}.
\end{abstract}

\begin{IEEEkeywords} 
Computational Photography, Retinal Images, Vessel Trees, Dataset, Denoising Diffusion Probabilistic Models, Segmentation.
\end{IEEEkeywords}
}

\maketitle

\IEEEraisesectionheading{
  \section{Introduction}\label{Introduction}
}
%
%
%
%

Retinal images play a vital role in diagnosing various diseases. By analyzing the retinal images, ophthalmologists can detect numerous health conditions related to eye, blood circulation, and brain-related disorders, e.g., retinal tear and detachment, diabetic and hypertensive retinopathy, papilledema, optic atrophy, microaneurysms, etc. These issues may lead to more severe health conditions, e.g., diabetic retinopathy is a consequence of diabetes, hypertensive retinopathy caused by hypertension, papilledema may lead to a brain tumor, meningitis, and stroke. These diseases can be detected by analyzing various characteristics of retinal vessel trees, such as length, width, curvature, and shape. Thanks to detecting these health conditions in advance medical treatment {is} greatly assisted. However, retinal image segmentation is a challenging task due to several factors. One of the most demanding difficulties {are} the fundus segmentation datasets that {are required for efficient training of learning-based models}. The models may often produce false positive results because there is a limited difference between vessel trees and background. Three popular retinal vessel segmentation datasets are available in the literature, including DRIVE, STARE, and CHASE DB1 \cite{staal2004ridge, hoover2000locating, fraz2012ensemble}. These datasets have only 40, 20, and 14 image pairs, respectively.

Many deep-learning-based retinal vessel segmentation models are reported in the literature. The authors focused on increasing the size and complexity of the models, leading to lower computational {efficiency}. For example, IterNet \cite{li2020iternet} is composed of one iteration of UNet \cite{ronneberger2015u} and several iterations of mini-UNet. Several learning-based methods \cite{guo2022novel, andreini2021two, niu2021explainable} have been proposed to increase the training data. The authors utilized GANs to synthesize retinal images. However, it {suffers from} several challenging problems, including difficulty in training as the models are susceptible to training parameters, inability to generate diverse data (mode collapse), vanishing gradient due to adversarial training, and non-convergence \cite{chen2021challenges}. 

In this work, we intend to develop a computationally efficient way of using Denoising Diffusion Probabilistic Model (DDPM) for retinal image generation and vessel segmentation. To do so, we propose a novel lightweight architecture with a new training technique leading to faster convergence than the original DDPM \cite{sohl2015deep} method. Our method has two main sub-models: (a) we perform a two-stage retinal image and corresponding vessel tree generation, (b) image super-resolution and segmentation. As shown in Fig. \ref{fig:workflow}, the model initially learns to generate vessel trees from the normal distribution before translating images from vessel trees to corresponding fundus images using our guided DDPM. In the third and fourth steps, we increase the resolution of generated images and train a biomedical segmentation UNet \cite{ronneberger2015u} to perform the retinal image segmentation task. Therefore, we propose a novel lightweight DDPM and a retinal segmentation dataset (ReTree). The proposed dataset has been evaluated using Frechet Inception Distance \cite{heusel2017gans} (FID). The ReTree dataset's efficiency is rated by comparing training UNet with real and our data and evaluating them with real testing images. We used the Jaccard similarity coefficient, Cohen's kappa, Matthew's Correlation Coefficient (MCC), F1-score, precision, recall, and accuracy to evaluate segmentation results. Super-resolution model was tested using Structural Similarity Index Measure \cite{wang2004image} (SSIM), Peak Signal-to-Noise Ratio (PSNR), and Binary Cross Entropy (BCE) loss. The main contributions of our work are summarized below. 

\begin{itemize}
    \item Develop a new Retinal Tree ReTree dataset for more efficient retinal segmentation using Denoising Diffusion Probabilistic Model (DDPM). Our dataset has 30,000 retinal images with corresponding vessel trees. 
    \item Propose a novel lightweight architecture of DDPM that leads to low computational cost, reducing the time and memory complexity of DDPM. In addition, it allows increasing the resolution of generated images by factors of $\times$2 and $\times$4.
    \item Design a new Repetitive Training Technique (RTT) for faster DDPM convergence. {Experimental results show that RTT is capable of increasing the quantitative metrics as well as reducing the training time.} In this training technique, the model gets re-trained if the loss in the current train step is higher than the global lowest train loss.
    \item Perform extensive quantitative and qualitative evaluation of the proposed model and dataset using three publicly available real retinal segmentation datasets.
\end{itemize}

The organization of this paper is as follows. In section \ref{Related Works}, retinal image generation, single image super-resolution, and segmentation methods available in the literature have been briefly reviewed. A detailed description of the proposed methodologies and implementation is provided in Section \ref{Methodology}. In Section \ref{Datasets}, we share information about utilized datasets, experimental setup, and model evaluation techniques. Section \ref{Results} demonstrates the results obtained by our method and the comparison with the state-of-the-art methods. Finally, we conclude the proposed method with future work plans in Section \ref{Conclusion}.

\section{Related Works}
\label{Related Works}

\subsection{Image Generation}
In several methods \cite{kim2022synthesizing, guo2022novel, andreini2021two, niu2021explainable, liu2019synthesizing, diaz2019retinal, yu2019retinal, appan2018retinal, guibas2017synthetic, costa2017end, costa2017towards}, authors developed deep learning models based on GANs \cite{goodfellow2020generative} to generate retinal images with corresponding label-maps. RetiGAN \cite{guo2022novel} developed a model with embedded Visual Geometry Group (VGG) network \cite{simonyan2014very} to improve the generated retinal images and trained it using vessel trees acquired from segmented fundus images using UNet \cite{ronneberger2015u}. {Kim et al. \cite{kim2022synthesizing} trained a GAN with large amount of training data for retinal image synthesis.} Andreini et al. \cite{andreini2021two} proposed a two-stage Progressively Growing GAN that generates semantic label-maps initially. Later, it performs the image-to-image translation from the vessel tree to the retinal image. Patho-GAN \cite{niu2021explainable} built a model that generates retinal images with symptoms related to diabetic retinopathy (DR) using vessel trees. Two pairs of images are fed into a discriminator model, and a pair of real and generated images are fed into the Perceptual network \cite{simonyan2014very}. Finally, a pair of images are passed to the DR detector to calculate the severity of it. {The authors of \cite{liu2019synthesizing}, utilized a combination of DCGAN and Wasserstein GAN to generate retinal images with diseases for their classification.} In another work \cite{diaz2019retinal}, authors utilized Deep Convolutional GAN (DCGAN) to generate retinal images for glaucoma assessment. Yu et al. \cite{yu2019retinal} used a pre-processing pipeline multiple-channels-multiple-landmarks (MCML) that produces images from a combination of vessel trees, optic disc, and cup images. First, they segment original fundus images and feed the output to a generator model. The generated result and a segmentation output are passed to the discriminator model with a real pair of the retinal and segmented images. On the other hand, Appan et al. \cite{appan2018retinal} utilized GAN to generate retinal images with lesions of various severity levels. They also developed a computer-aided diagnosis system to detect hemorrhage, which was trained using previously generated data. Costa et al. \cite{costa2017end} utilized GAN architecture to generate vessel trees and then fundus images from these vessel trees. First, they built two autoencoders using adversarial learning to generate vessel trees and their corresponding retinal images. In another work \cite{costa2017towards}, they developed a retinal GAN to generate retinal images from related vessel trees. In the available literature, authors utilized only GAN-based methods for retinal image synthesis. However, GANs are proven to have such issues as vanishing gradients, mode collapse, and non-convergence. Therefore, this work focused on solving these issues by utilizing a new DDPM-based technique that has achieved state-of-the-art performance in generating retinal images. 

{The first original generative DDPM was proposed by Sohl-Dickstein et al. \cite{sohl2015deep} and Ho et al. \cite{ho2020denoising}. Song et al. \cite{song2020denoising} proposed denoising diffusion implicit models (DDIMs), a new class of diffusion models that uses non-Markovian diffusion processes. In another work \cite{choi2021ilvr}, the authors proposed a DDPM with a novel Iterative Latent Variable Refinement (ILVR), that conditions the generative process to synthesize high-quality images based with a given reference image. In this work, we compare the proposed model with Improved DDPM \cite{nichol2021improved} (IDDPM), which is the current state-of-the-art method in image generation task. It utilizes Vision Transformer (ViT) encoder blocks with Multi-Head Self-Attention after each down- and up-sample block. This approach reshapes the features by a factor of 2 using 4 down- and up-sample blocks. Due to the architecture of IDDPM \cite{nichol2021improved}, it is impossible to generate images with resolution other than 64$\times$64, unlike with the proposed model. Therefore, we up-scale the generated images using Enhanced Super-Resolution GAN (ESRGAN) proposed by Wang et al. \cite{wang2018esrgan}. ESRGAN \cite{wang2018esrgan} is based on SRGAN \cite{ledig2017photo}, that was the first GAN used in single-image super-resolution. ESRGAN \cite{wang2018esrgan} utilizes Residual-in-Residual Dense Block (RRDB) as the main building block of the model.} Recently, a new work has been proposed \cite{alimanov2023hybrid} for retinal image super-resolution using ViT and Convolutional Neural Network (CNN). Qiu et al. \cite{qiu2022improved} proposed Improved GAN with a novel residual attention block for a more accurate generation.

\begin{figure*}[ht!]
    \centering
    \includegraphics[width=0.85\textwidth]{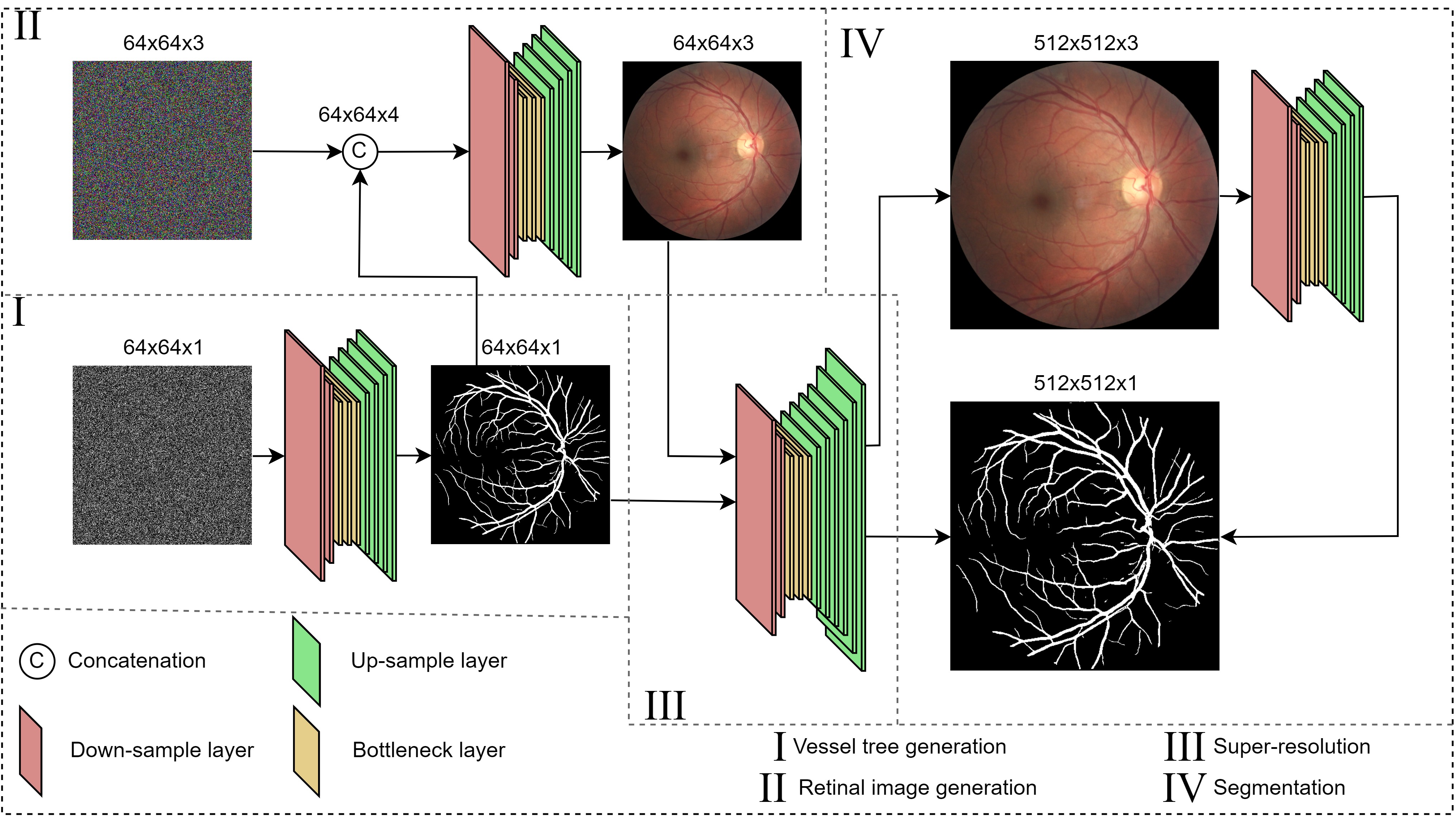}
    \caption{Workflow of the proposed framework. In the first (I) step, we train DDPM to generate vessel trees from noise. Next (II), another DDPM learns to generate retinal images from noise and vessel semantic label-map. In the third (III) step, we performed super-resolution of generated vessel trees and retinal images. Finally (IV), the biomedical segmentation model is utilized to segment generated up-sampled retinal images for validation.}
    \label{fig:workflow}
\end{figure*}

\subsection{{Retinal} Image Segmentation}
Li et al. \cite{li2020iternet} proposed a UNet-based \cite{ronneberger2015u} IterNet that adopts several iterations of a mini-UNet. In each iteration, the features from previous steps are shared using short and long skip connections. In Attention Guided
Network (AG-Net) \cite{zhang2019attention}, the authors proposed a method based on M-Net \cite{fu2018joint} with attention-guided filter \cite{he2012guided} that transfers structural information from low-level feature maps to high-level. Jiang et al. \cite{jiang2021multi} proposed Multi-Scale and Multi-Branch Network (MSMB-Net) for retinal image segmentation, utilizing atrous convolutions and skip connections for more efficient feature extraction. Li et al. \cite{li2021ta} in the Triple Attention Network (TA-Net) embedded a channel with a self-attention encoder and spatial attention up-sampling blocks to improve the representation of target features and to train the model to pay more attention to the essential pixels. Zhang et al. \cite{zhang2021pyramid} utilized pyramid UNet with pyramid-scale aggregation blocks for feature aggregation at all levels. Recently, a few methods \cite{baranchuk2021label, amit2021segdiff} utilized the DDPM in image segmentation. For example, Amit et al. \cite{amit2021segdiff} employed DDPM for natural image segmentation. After passing them to convolutional encoders, the authors used feature addition instead of concatenation for input images. {In another work \cite{alimanov2022retinal1}, UNet with Convolutional Block Attention Module (CBAM-UNet) has been proposed for more accurate retinal vessel segmentation.} 


\section{Methodology}
\label{Methodology}

\subsection{Background}
The DDPM uses two stages to generate samples: forward and backward processes. In the forward diffusion process, we gradually add Gaussian noise to the input image through a series of diffusion time steps $T$. In the backward process, the model tries to denoise the input at the given diffusion step $t$. In the forward diffusion, at each diffusion step $t$, we sample $x_t$ from data distribution $q(x)$ ($x_t \in q(x)$). The noise is added to the input variable $x_{t-1}$ using the variance schedule $\beta_t$ to acquire $x_t$ with $q(x_t|x_{t-1})$. To do so, we use the following formula.
\begin{equation}\label{eq:1}
    q(x_t|x_{t-1}) = \mathcal{N}(x_t; \mu_t = \sqrt{1 - \beta_t}x_{t-1}, \Sigma_t = \beta_tI)
\end{equation}
where $\mathcal{N}$ is the Gaussian distribution, $\mu_t$ is the mean, $\Sigma_t$ is the standard deviation, a diagonal matrix of $\beta_t$ values because $I$ is the identity matrix.

With equation \ref{eq:1}, it is possible to apply noise to $x_0$ in a tractable way (1:$T$) to get $x_T$. This posterior probability is defined as:
\begin{equation}
    q(x_{1:T}|x_0) = \prod_{t=1}^{T}q(x_t|x_{t-1})
\end{equation}
However, with these equations, to sample $x_T$, we would have to apply {$q(x_{1:T}|x_0)$ a total of $T$ number of times} repeatedly. Therefore, a reparametrization trick was used to sample $x$ at any time step $t$, which is defined in the following formula:
\begin{equation} \label{eq:2}
    \begin{split}
        x_t & = \sqrt{1-\beta_t}x_{t-1} + \sqrt{\beta_t}\epsilon_{t-1} \\
            & = \sqrt{\alpha_t}x_{t-2} + \sqrt{1-\alpha_t}\epsilon_{t-2} \\
            & = \dots \\
            & = \sqrt{\Bar{\alpha}_t}x_{0} + \sqrt{1-\Bar{\alpha}_t}\epsilon_{0}
    \end{split}
\end{equation}
where $\alpha_t$ = $1-\beta_t$, $\Bar{\alpha}_t$ = $\prod_{s=0}^{t}\alpha_s$ and noise $\epsilon_0, \dots, \epsilon_{t-2}, \epsilon_{t-1} \sim \mathcal{N}(0, I)$. 

This reparametrization trick allows us to precompute $\alpha_t$ and $\Bar{\alpha}_t$ for any diffusion step $t$ to sample $x_t$ in one operation, to produce $x_t$ we can use the following distribution:
\begin{equation}
    x_t \sim q(x_t|x_0) = \mathcal{N}(x_t; \sqrt{\Bar{\alpha}_t}x_0, (1-\Bar{\alpha}_t)I)
\end{equation}
In our work, $\beta_1$ and $\beta_T$ are 0.0001 and 0.02, respectively. In addition, we utilized cosine noise scheduler \cite{nichol2021improved} instead of the linear one used in the original DDPM work \cite{sohl2015deep} as it has shown to lead to faster convergence.

\begin{figure*}[ht!]
    \centering
    \includegraphics[width=0.9\textwidth]{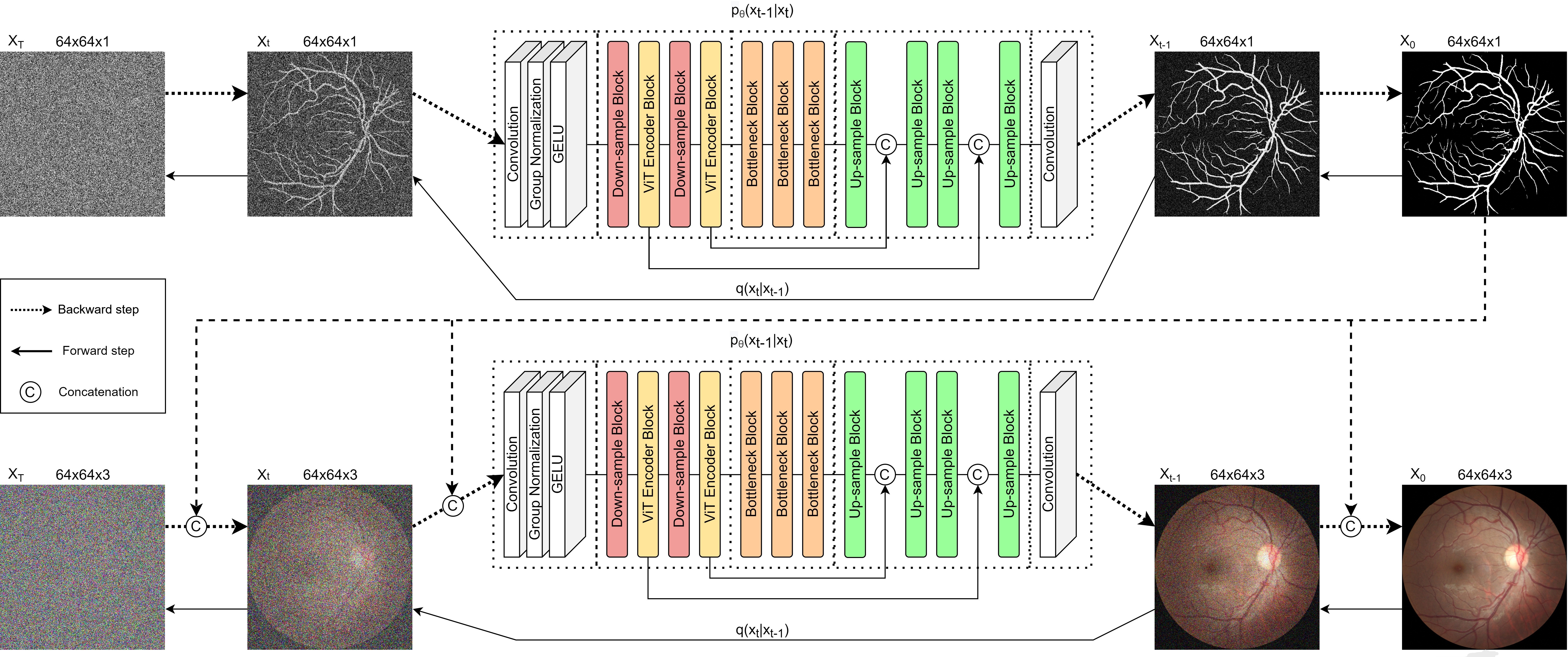}
    \caption{The general architecture of the proposed DDPMs. The first model generates semantic label maps from noise in a given number of time steps ($T$). At each time step $t$, the model predicts output from time steps $t-1$. In the second step, the model takes RGB noise and {the} output of the first model to generate fundus images. The image containing vessel trees is concatenated with noisy input at each diffusion step $t$. Each DDPM model includes initial, down-sample, ViT encoder, bottleneck, and up-sample blocks.}
    \label{fig:ddpms}
\end{figure*}

In the reverse diffusion process, the model is trying to learn distribution $p_{\theta}(x_{t-1}|x_t)$, which is an approximation of original distribution $q(x_{t-1}|x_t)$. In this process, a deep learning method is utilized to gradually remove noise at each diffusion step from $T$ to 1. This process can be defined as follows:
\begin{equation}
    p_{\theta}(x_{t-1}|x_t) = \mathcal{N}(x_{t-1}; \mu_{\theta}(x_t, t), \Sigma_{\theta}(x_t, t))
\end{equation}
This reverse formula is applied to all diffusion steps to compute the trajectory $p_{\theta}(x_{0:T})$ using equation \ref{eq:3}:
\begin{equation} \label{eq:3}
    p_{\theta}(x_{0:T}) = p_{\theta}(x_T)\prod_{t=1}^{T}q(x_{t-1}|x_t)
\end{equation}
However, the output of a DDPM is not an image itself. The model takes the noisy image as input and predicts the noise. To sample an image from complete noise input $x_T$, we use the following formula \ref{eq:4}:
\begin{equation} \label{eq:4}
    x_{t-1} = \frac{x_t - \frac{1 - \alpha_t}{\sqrt{1 - \Bar{\alpha_t}}}\epsilon_{\theta}(x_t, t)}{\sqrt{\alpha_t}} + \sqrt{\beta_t}z
\end{equation}
where $\epsilon_{\theta}(x_t, t)$ is predicted noise by DDPM and if $t$ is not equal to 1 then $z \sim \mathcal{N}(0, I)$, otherwise $z$ is 0. Equation \ref{eq:4} is repeated $T$ times, from $t=T$ until $t=1$ to acquire $x_0$.

\subsection{Overview of Proposed Framework}
Our work focused on building a framework for retinal images with vessel tree generation, super-resolution, and segmentation. The workflow of the proposed method is shown in Fig. \ref{fig:workflow}, which has four major steps. Firstly, we develop the denoising diffusion probabilistic model (DDPM) to generate semantic label maps from noise with one channel. Secondly, we applied an example-based generation technique for guided training of another DDPM to generate colorful fundus images from given vessel trees. In this step, the model synthesizes retinal images using noise with red, green, and blue (RGB) channels concatenated with a grayscale vessel tree; concatenation is performed at each diffusion time step $t$ as shown in Fig. \ref{fig:ddpms}. After generating retinal images and vessel maps, the resulting images require up-scaling to improve the image quality for efficient image segmentation. The output of DDPMs is 64$\times$64-pixel images. 
Therefore, we performed super-resolution of retinal images up to 512$\times$512 pixels using ESRGAN \cite{wang2018esrgan} with modifications for a specific task (RGB and binary images super-resolution). To up-sample the retinal images, we added SSIM loss in equation \ref{eq:ssim} during training of the ESRGAN model for preserving the image structure. To perform super-resolution of vessel maps, we had to remove the Perceptual loss computed using VGG-19 \cite{simonyan2014very} because it requires input images with 3 channels. However, we included a BCE loss, shown in equation \ref{eq:bce} to evaluate the performance of binary image super-resolution. In the final step, we performed vessel segmentation of the proposed dataset using UNet \cite{ronneberger2015u}. 

\subsection{ReTree Image Generation Models}
\begin{figure}[ht!]
    \centering
    \includegraphics[width=0.45\textwidth]{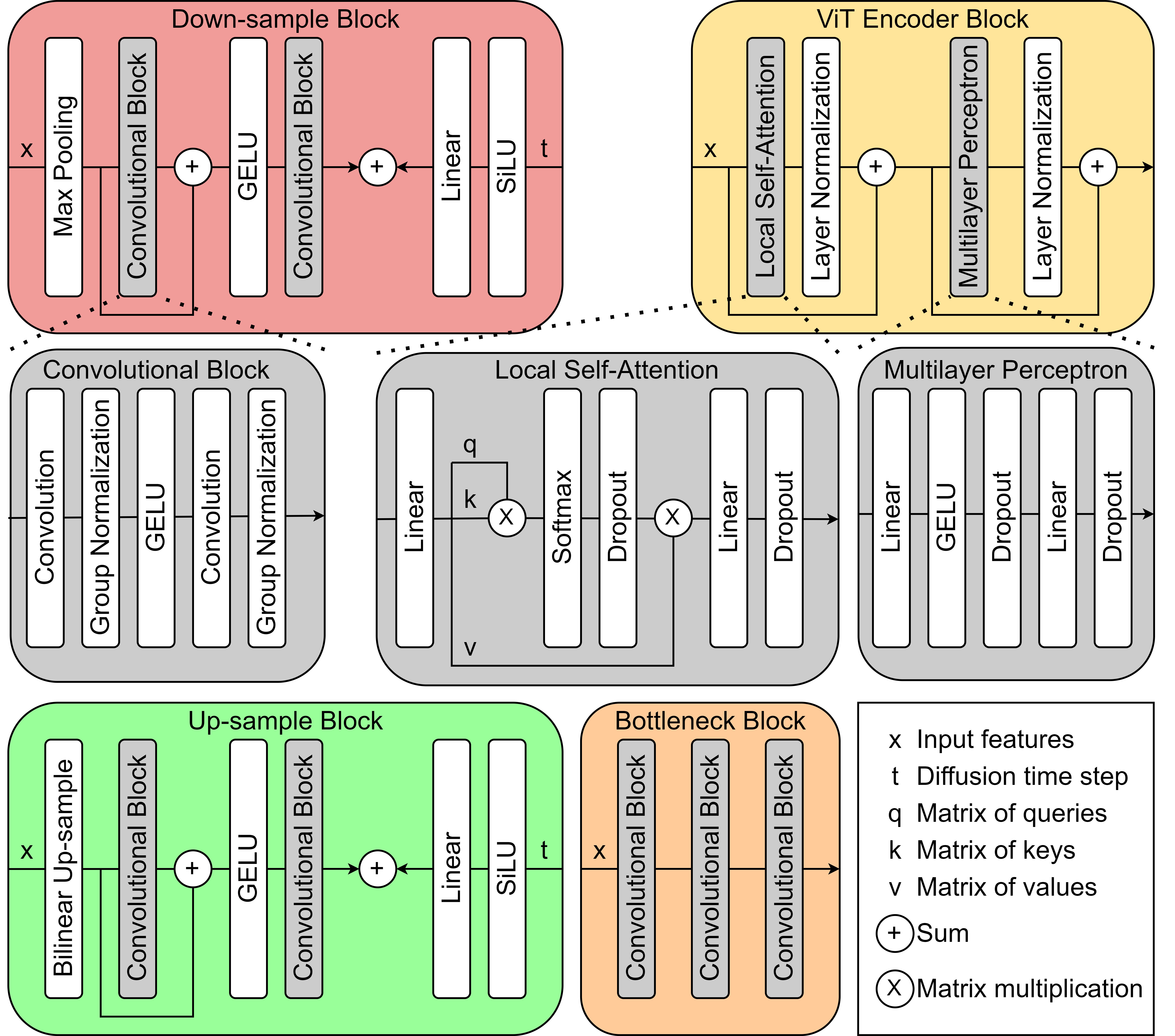}
    \caption{The detailed architecture of the proposed DDPM. It consists of down-sampling, ViT encoder, up-sampling, and bottleneck blocks.}
    \label{fig:ddpm}
\end{figure}

Retinal image generation consists of two stages: retinal vessel generation (1) and retinal image synthesis (2). In both tasks, we utilized DDPM, the architectures of proposed DDPMs are presented in Fig. \ref{fig:ddpms} and in Fig. \ref{fig:ddpm}. Both models comprise initial down-sample, ViT encoder, bottleneck, and up-sample blocks. The initial block includes 2D convolution with channel number, kernel size, and padding equal to 128, 3, and 1, respectively. After that, the input $x_t$ is normalized and activated using group normalization and Gaussian Error Linear Unit (GELU) \cite{hendrycks2016gaussian}. Next, $x_t$ and the diffusion step are sent to one down-sample block. The input $x_t$ is down-scaled using one max pooling layer with a window size of 4 and two convolutional blocks. Each block is built with convolution, group normalization, GELU, convolution, and group normalization layers. Also, diffusion step $t$ is activated using Sigmoid Linear Unit \cite{elfwing2018sigmoid} (SiLU) fed into a linear layer. The resulting embeddings are added to $x_t$ and processed by ViT encoder block with locality self-attention (LSA) \cite{lee2021vision}. The procedure with down-sampling and ViT blocks are repeated. The bottleneck is composed of three consecutive convolutional blocks. The up-scaling is performed using the bilinear up-sampling with a factor of 2 and two convolutional blocks with a residual connection. Since down-sampling was performed twice with a factor of 4, the model includes four up-sampling blocks. During up-scaling $x_t$, we need to add time embeddings acquired from $t$ in the same way as in down-sampling. Finally, $x_t$ is fed into the last convolutional layer with a kernel size of 1.

The LSA was first proposed as a solution for small datasets by Lee et al. \cite{lee2021vision}. Nonetheless, when trained on medium-sized datasets, it has also outperformed multi-head self-attention. Thanks to learnable temperature scaling. We employ LSA to train our model with the mid-sized EyeQ dataset effectively. The queries and keys matrices are multiplied in LSA to obtain $R_{i, j}$. It uses diagonal masking to improve the attention process. The masking is shown in equation \ref{eq:mask}. The diagonal elements are set to $-\infty$. Using equation \ref{eq:lsa}, we get attention values for $R$. 
\begin{equation} \label{eq:mask}
    R^{Masked}_{i, j} = 
    \begin{cases}
    R_{i, j} & i\neq j \\
    - \infty & i = j
    \end{cases}
\end{equation}
{where $R_{i, j}$ represents the components of the dot product between the matrix of queries and transposed matrix of keys $R = q \times k^T$.} 
\begin{equation} \label{eq:lsa}
    LSA = Softmax(\frac{R^{Masked}}{\tau})
\end{equation}
{where $\tau$ is learnable temperature scaling, another feature of LSA that helps the softmax calculate its temperature during training.}

\subsection{{Repetitive Training Technique}}
{The proposed Repetitive Training Technique is a simple yet effective approach that experimentally demonstrates a reduction in training time and an improve in quantitative results. This technique aims to enhance the robustness of the proposed model against randomness and outliers. Specifically, when outlier noise is introduced, the backward diffusion process may yield unrealistic results. Therefore, re-training the model with the same data is beneficial until the loss decreases. By employing this technique, we observed a decrease in the number of epochs and total training time, along with a reduction in the generation of unrealistic images, leading to a lower FID score. In our experiments, the maximum number of repetitions is set to 5. If the current loss value surpasses the global minimum average loss across all epochs, the model is re-trained with the current data, repeating this process up to 5 times.}

\subsection{Loss Functions}
DDPMs were trained with a combination of 2 loss functions which is the sum of L1, or Mean Absolute Error (MAE), and Mean Squared Error (MSE) between predicted noise and applied noise, as shown below:
\begin{equation}
    Gen_{Loss}(\epsilon_{\theta}(x_t, t), z) = \frac{1}{N}\sum_{i=1}^{N}(|\epsilon_{\theta} - z| + (\epsilon_{\theta} - z)^2)
\end{equation}
where $N$ is the total number of image pixels, $\epsilon_{\theta}$ is the predicted noise and $z$ is the actual applied noise. {The combined loss produced more accurate results and performance.}

Super-resolution model was trained with a combination of 5 loss functions, such as L1, Structural Similarity Index Measure (SSIM), adversarial loss, Binary Cross Entropy (BCE), and Perceptual loss. SSIM computes the structural similarity between two images, up-sampled $X$ and ground truth $Y$, using the following equation:
\begin{equation}\label{eq:5}
    SSIM(X, Y) = \frac{(2\mu_{X}\mu_{Y} + C_1)(2\sigma_{XY} + C_2)}{(\mu^{2}_{X} + \mu^{2}_{Y} + C_1)(\sigma^{2}_{X} + \sigma^{2}_{Y} + C_2)}
\end{equation}
{where $\mu$ is the mean, $\sigma$ is the variance, and $C_1$ and $C_2$ are variables that stabilize the division.}

Using equation \ref{eq:5} we can define the SSIM loss function in the following way:
\begin{equation} \label{eq:ssim}
    SSIM_{Loss}(X, Y) = \frac{1 - SSIM(X, Y)}{2}
\end{equation}

Adversarial loss is calculated using the output of a discriminator model, which is responsible for classifying the input images as real or fake. {The discriminator model used to filter the generated images contains an initial Convolutional layer with LeakyReLU activation function, 3 Convolutional down-sampling blocks with Instance Normalization and LeakyReLU layers along with the ViT encoder blocks, and a final Convolutional layer. Each down-sampling block down-scales the input with a factor of 2. The number of kernels is 4, and the number of strides is 2, 1, 2, 2, and 1, respectively. The output prediction is activated using the Sigmoid function.} The discriminator loss is defined below:

\begin{equation}
\begin{split}
    Adv_{Loss}(Y, X) = &-\mathbb{E}_{Y}[log(1 - D(Y, X))] \\
                               &-\mathbb{E}_{X}[log(D(Y, X))]
\end{split}
\end{equation}
where $D(x_r, x_f)$ is the prediction given by the discriminator model.

Perceptual loss, which measures the distance between activated features, is implemented using VGG19-54 \cite{simonyan2014very}. The distance is minimized using the MSE loss function.

The segmentation model is trained with Binary Cross-Entropy (BCE) loss. It is done using the following equation:
\begin{equation}\label{eq:bce}
\begin{split}
    BCE_{Loss} = &- \frac{1}{N}\sum_{i=1}^{N}(x_{i_{t-1}}log(x_{i_{t-1}}^{\theta}) \\
    &+ (1-x_{i_{t-1}})log(1-x_{i_{t-1}}^{\theta}))
\end{split}
\end{equation}
where $N$ represents the number of pixels in the image.

\section{Datasets and Experimental Setup}
\label{Datasets}
\subsection{Datasets}
In this work, we utilized a total of 4 retinal datasets for image generation, super-resolution, and segmentation tasks. At first, we trained UNet \cite{ronneberger2015u} with DRIVE \cite{staal2004ridge}, STARE \cite{hoover2000locating}, and CHASE DB1 \cite{fraz2012ensemble} combined {to segment retinal images}. Next, we used this model to segment the EyeQ dataset \cite{fu2019evaluation} that comprises 28,792 512$\times$512 pixels RGB retinal images belonging to 3 classes according to their quality. The classes include "Good", "Usable", and "Bad". There are 16,818" Good", 6,434" Usable," and 5,538 "Bad" images; {additionally, the dataset contains images with diabetic retinopathy, glaucoma, and age-related macular degeneration}. In step 1, the DDPM is trained with segmented images from the EyeQ dataset that belong to the "Good" quality class. In the next step, the same segmented images and corresponding RGB fundus images were utilized for training the second DDPM to generate retinal images. In the next step, Enhanced Super-Resolution GAN (ESRGAN) \cite{wang2018esrgan} was trained using the same image pairs from the "Good" quality EyeQ dataset. Using these DDPMs and ESRGAN, we generated and up-scaled pairs of images for the ReTree dataset. Finally, the segmentation UNet \cite{ronneberger2015u} was trained with generated image pairs and tested using DRIVE, STARE, and CHASE DB1 to validate the efficiency of the proposed dataset. Table \ref{table:datasets} shows detailed information about the used datasets for the experiment.

\setlength{\extrarowheight}{2pt}

\begin{table}[htb]
\caption{Details of datasets used for the experiment}
\label{table:datasets}
\centering
\scalebox{0.8}{
 \begin{tabular}{| c  c  c  c| } 
 \hline
 \textbf{Dataset Name} & \textbf{No. of Images} & \textbf{Resolution} & \textbf{Class} \\ [0.5ex]
 \hline\hline
 \multirow{3}{*}{\specialcell{EyeQ \cite{fu2019evaluation}}} & 16,818 & 512$\times$512 & Good \\ [0.5ex] 
 & 6,434 & 512$\times$512 & Usable \\ [0.5ex]
 & 5,538 & 512$\times$512 & Bad \\ [0.5ex]
 \hline
 \multirow{2}{*}{\specialcell{DRIVE \cite{staal2004ridge}}} & 20 & 584$\times$565 & Train \\ [0.5ex] 
 & 20 & 584$\times$565 & Test \\ [0.5ex] 
 \hline
 \multirow{2}{*}{\specialcell{STARE \cite{hoover2000locating}}} & 10 & 700$\times$605 & Train \\ [0.5ex]
 & 10 & 700$\times$605 & Test \\ [0.5ex]
 \hline
 \multirow{2}{*}{\specialcell{CHASE DB1 \cite{fraz2012ensemble}}} & 14 & 999$\times$960 & Train \\ [0.5ex]
 & 14 & 999$\times$960 & Test \\ [0.5ex]
 \hline
 \end{tabular}}
\end{table}


\begin{figure*}[ht!]
    \centering
    \begin{minipage}{1\textwidth}
    \hspace{2.9cm} Real images from EyeQ \cite{fu2019evaluation} \hspace{3.2cm} Generated images from ReTree
    \end{minipage}
    \includegraphics[width=0.145\textwidth]{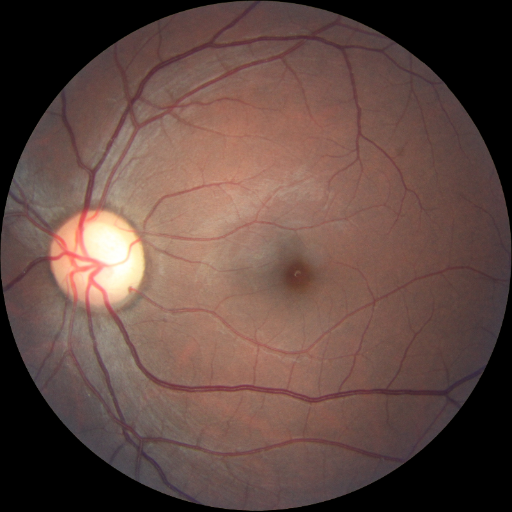}
    \includegraphics[width=0.145\textwidth]{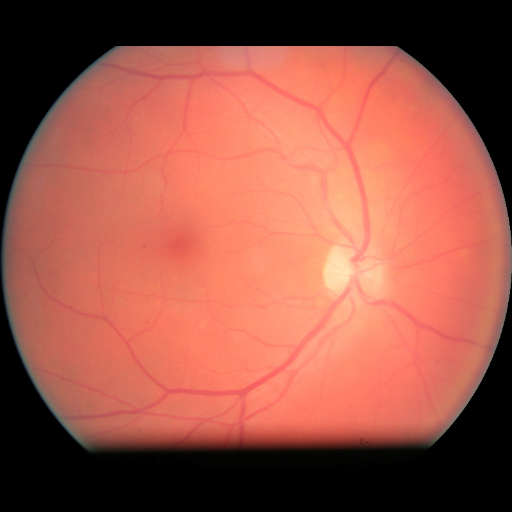}
    \includegraphics[width=0.145\textwidth]{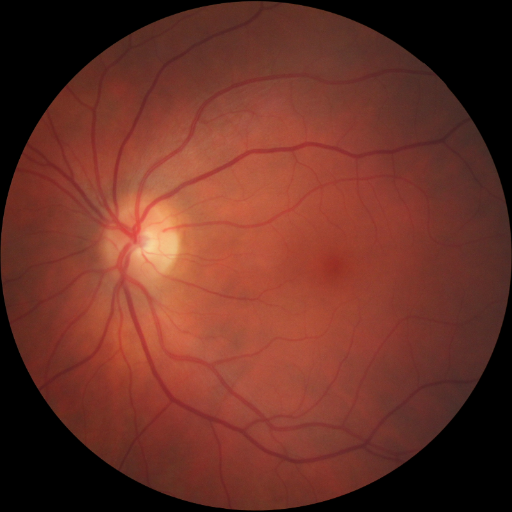}
    \hspace{0.05cm}
    \includegraphics[width=0.145\textwidth]{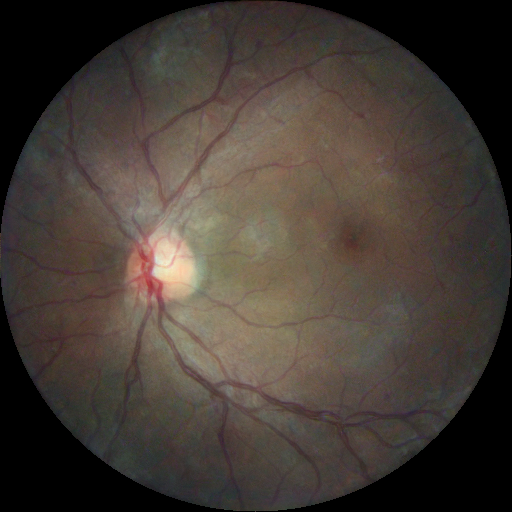}
    \includegraphics[width=0.145\textwidth]{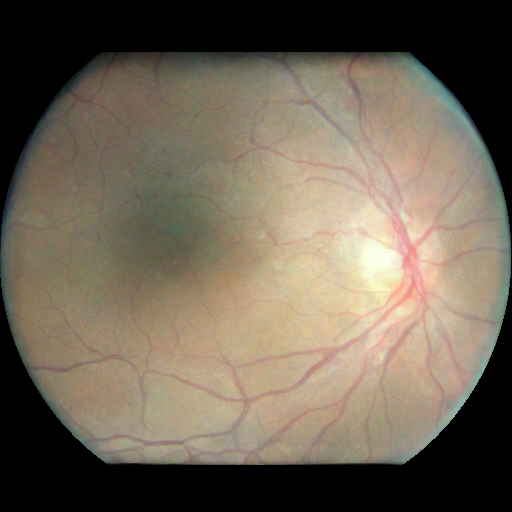}
    \includegraphics[width=0.145\textwidth]{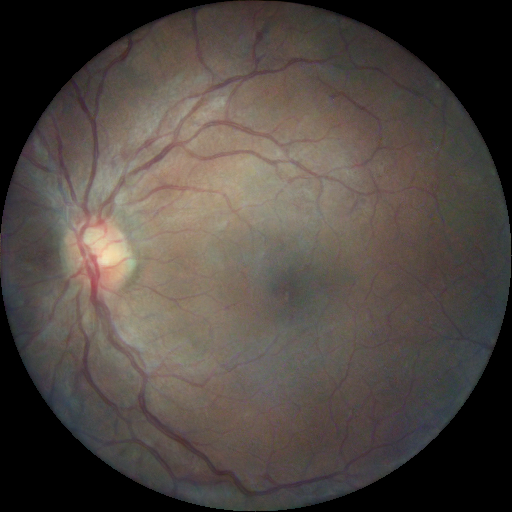}
    
    \vspace{0.05cm}

    \includegraphics[width=0.145\textwidth]{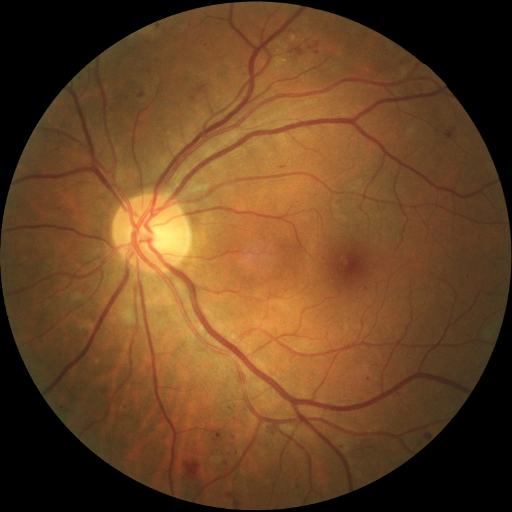}
    \includegraphics[width=0.145\textwidth]{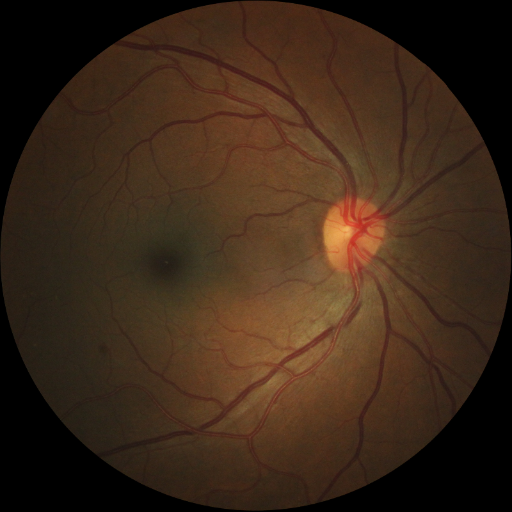}
    \includegraphics[width=0.145\textwidth]{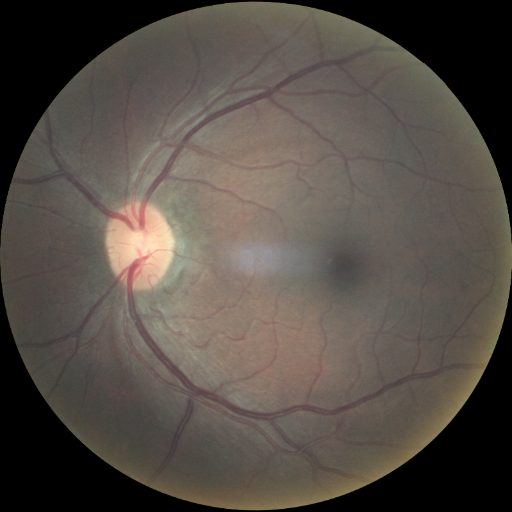}
    \hspace{0.05cm}
    \includegraphics[width=0.145\textwidth]{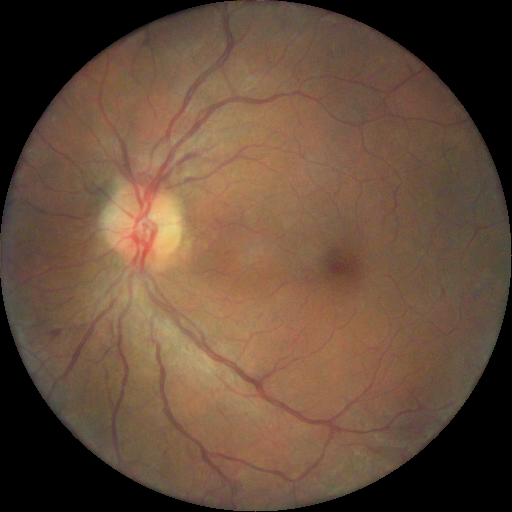}
    \includegraphics[width=0.145\textwidth]{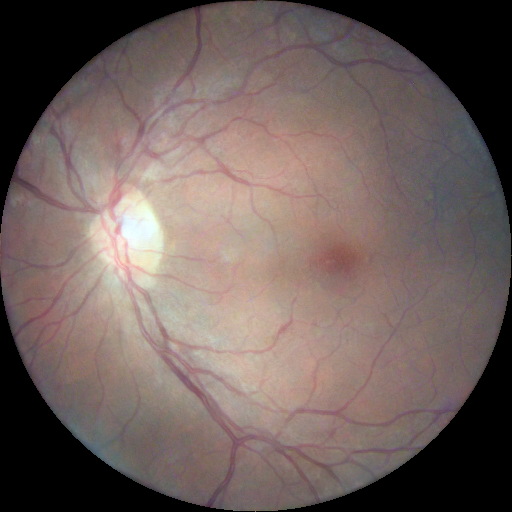}
    \includegraphics[width=0.145\textwidth]{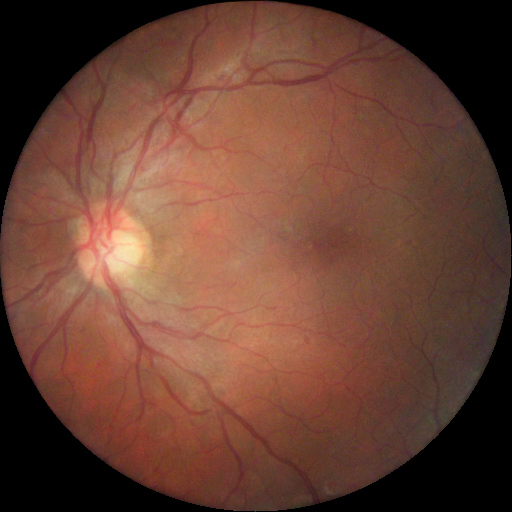}
    
    \vspace{0.05cm}

    \includegraphics[width=0.145\textwidth]{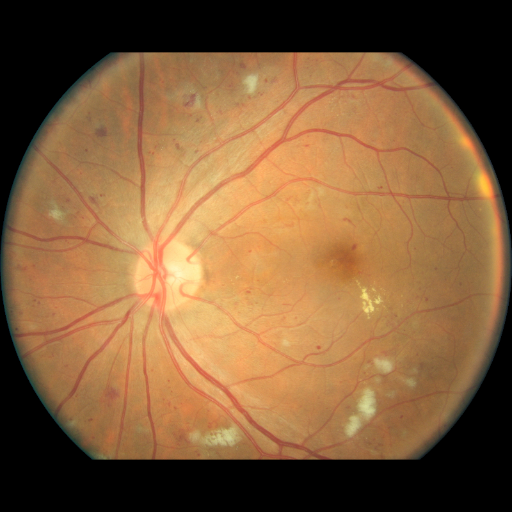}
    \includegraphics[width=0.145\textwidth]{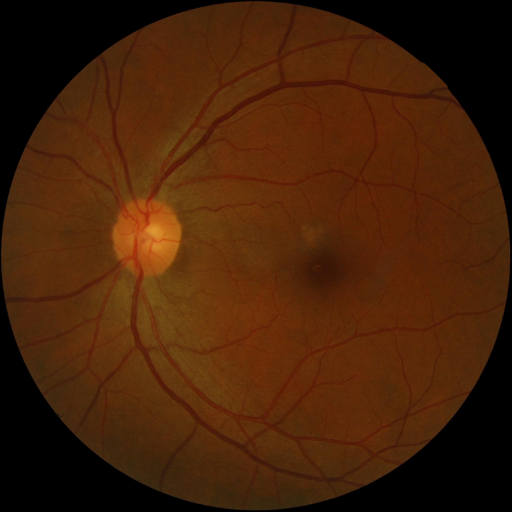}
    \includegraphics[width=0.145\textwidth]{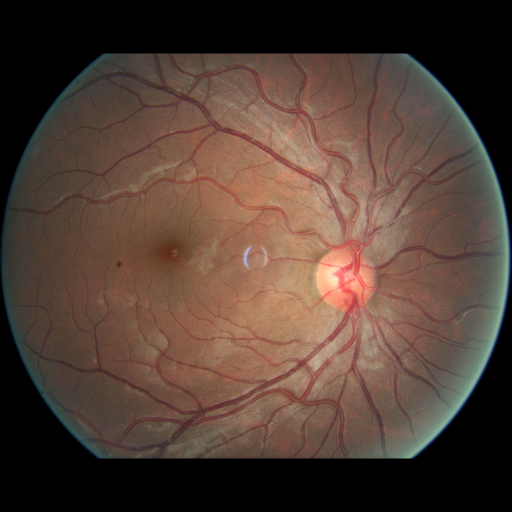}
    \hspace{0.05cm}
    \includegraphics[width=0.145\textwidth]{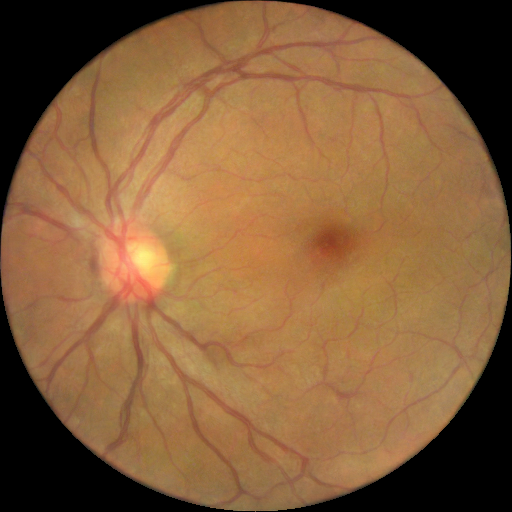}
    \includegraphics[width=0.145\textwidth]{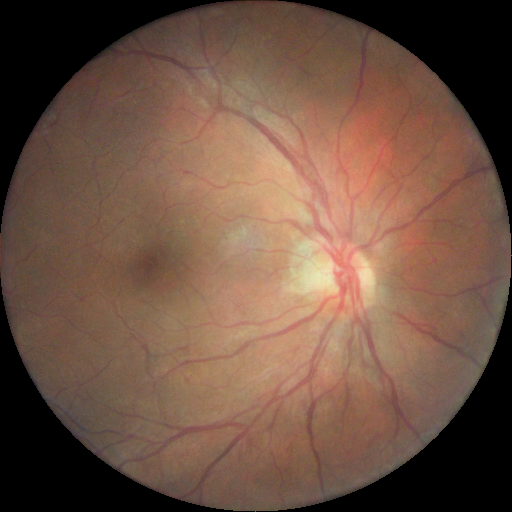}
    \includegraphics[width=0.145\textwidth]{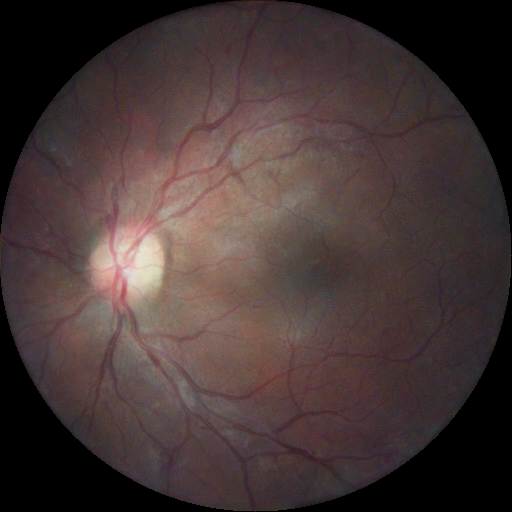}
    \caption{The generated results of the proposed method (on the right) along with real images from the EyeQ \cite{fu2019evaluation} dataset (on the left). The generated images by our ReTree method have similar qualities as real retinal images.}
    \label{fig:qualitative1}
\end{figure*}

\subsection{Experimental Setup}
The model has been implemented using the PyTorch framework. The hardware configurations are an Intel Core i7-10700f CPU, 32 GB RAM, and an NVIDIA GeForce RTX 2080 SUPER 8 GB GPU. In our experiments, we used the Adam optimization function, batch size of 32, and learning rate of $10^{-5}$. 

\subsection{Evaluation Metrics}
We computed the FID score \cite{heusel2017gans} to evaluate the quality and select realistic synthetic images. Additionally, we trained classification networks for generated vessel trees and retinal images using original and generated pictures combined. The output of these classification networks is a binary number, where 0 is a fake, and 1 is an actual image. The threshold was experimentally set to 0.8, {meaning the prediction should be above} 0.8, to select the most likely results for the ReTree dataset. The FID score is an evaluation metric for generative models. It uses the following formula for calculating the distance:
\begin{equation}
    FID = |\mu_1 - \mu_2| + Tr(\sigma_1 + \sigma_2 - 2*\sqrt{\sigma_1*\sigma_2})
\end{equation}
where $\mu_1$ and $\sigma_1$ refer to the mean and covariance of the generated images, while $\mu_1$ and $\sigma_1$ refer to the mean and covariance of the real images.

Super-resolution network has been tested using SSIM (\ref{eq:5}) and PSNR on the EyeQ dataset. The PSNR between generated image $X$ and real image $Y$ is computed using equation \ref{eq:6}.
\begin{equation}\label{eq:6}
    PSNR(X, Y) = 10 \log_{10} \left( \frac{MAX_X^2}{MSE} \right)
\end{equation}
where $MAX_X^2$ is the maximum value of image pixels.

A segmentation step was validated using statistical metrics, such as the Jaccard similarity index, precision, recall, F1-score, and accuracy. The Jaccard index calculates the similarity between two sets and is defined in equation \ref{eq:7}.
\begin{equation} \label{eq:7}
    J(X, Y) = \frac{|X \cap Y|}{|X \cup Y|}
\end{equation}
F1-score, recall, precision, and accuracy are calculated in the following ways:
\begin{align} 
    Recall &= \frac{TP}{TP + FN} \\
    Precision &= \frac{TP}{TP + FP} \\
    F1 &= \frac{2 * Precision * Recall}{Precision + Recall} \\
    Accuracy &= \frac{TP + TN}{TP + TN + FP + FN}
\end{align}
TP is true positive values, TN means true negative results, FP is false positive outcomes, and FN is false negative outputs.

In addition, we evaluated segmentation results using MCC and kappa because they are more accurate at assessing the classification methods since they do not suffer from sensitivity to class imbalance and asymmetricity, unlike precision, recall, F1, and accuracy. They are calculated using the following equation \ref{eq:mcc} and equation \ref{eq:kappa}.
\begin{equation} \label{eq:mcc}
    \footnotesize{MCC = \frac{TP*TN-FP*FN}{\sqrt{(TP+FP)(TP+FN)(TN+FN)(TN+FN)}}}
\end{equation}
\begin{equation} \label{eq:kappa}
    kappa = \frac{Accuracy - Accuracy^{R}}{1 - Accuracy^{R}}
\end{equation}
{where random accuracy $Accuracy^{R}$ is calculated in the following way:}
\begin{align} 
    Accuracy^{R} &= p_1 \times p_2 + (1 - p_1) \times (1 - p_2) \\ 
    p_1 &= \frac{TP + FN}{TP + TN + FP + FN} \\ 
    p_2 &= \frac{TP + FP}{TP + TN + FP + FN}
\end{align}

\section{Results and Discussion}
\label{Results}

\subsection{ReTree dataset}
We are proposing a retinal segmentation dataset (ReTree), which consists of images belonging to 2 classes: colored fundus images and corresponding binary vessel maps. The total number of images in each class is 30,000. The dataset is split into the train, validation, and test sets with 28,800, 200, and 1,000 images, respectively. The image resolution is 512$\times$512 pixels. 
The proposed dataset has been extensively evaluated using quantitative and qualitative measures. To validate the effectiveness of our dataset, we trained the biomedical segmentation UNet \cite{ronneberger2015u} model with it and tested the model with real images. The results indicate the improvement over training with solely the real images since the number of real images needs to be significantly increased to train the deep learning model efficiently. Fig. \ref{fig:qualitative1} presents the qualitative comparison with EyeQ \cite{fu2019evaluation}, where we can see the similarity and realism of generated images, having similar qualities such as vessel structure, accurate disk/cup region, and realistic color. 


\begin{figure}[h!]
    \centering
    \begin{minipage}{1\textwidth}
    \hspace{1.8cm} GAN \cite{costa2017end} \hspace{2.5cm} Ours
    \end{minipage}\vspace{0.07cm}
    \includegraphics[width=0.1\textwidth]{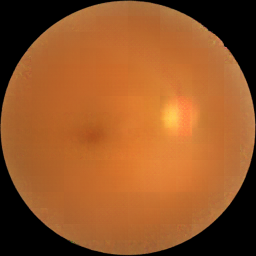}
    \includegraphics[width=0.1\textwidth]{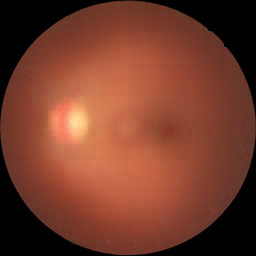}
    \hspace{0.01cm}
    \includegraphics[width=0.1\textwidth]{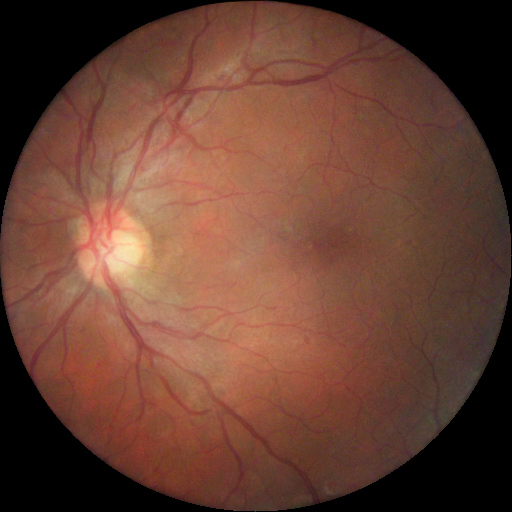}
    \includegraphics[width=0.1\textwidth]{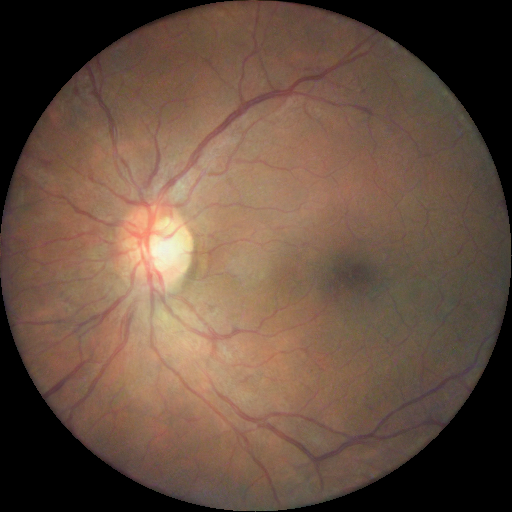}

    \vspace{0.05cm}

    \includegraphics[width=0.1\textwidth]{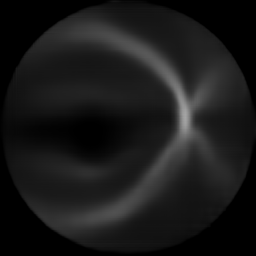}
    \includegraphics[width=0.1\textwidth]{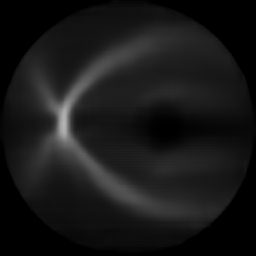}
    \hspace{0.01cm}
    \includegraphics[width=0.1\textwidth]{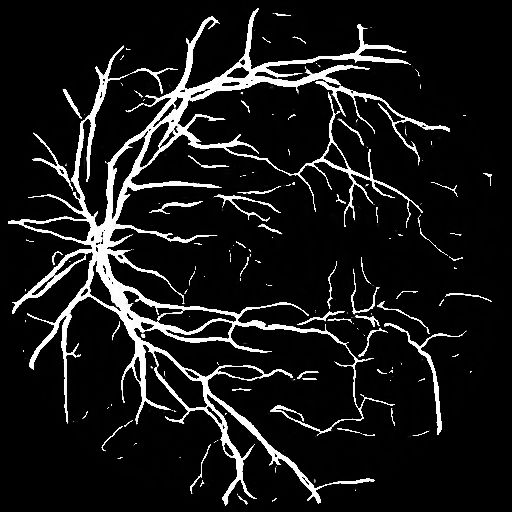}
    \includegraphics[width=0.1\textwidth]{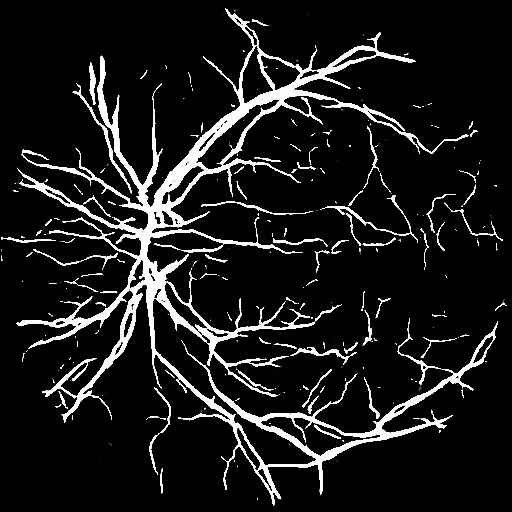}

    \vspace{0.05cm}
    
    \includegraphics[width=0.1\textwidth]{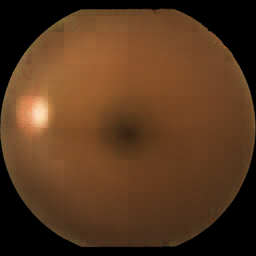}
    \includegraphics[width=0.1\textwidth]{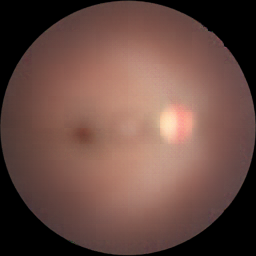}
    \hspace{0.01cm}
    \includegraphics[width=0.1\textwidth]{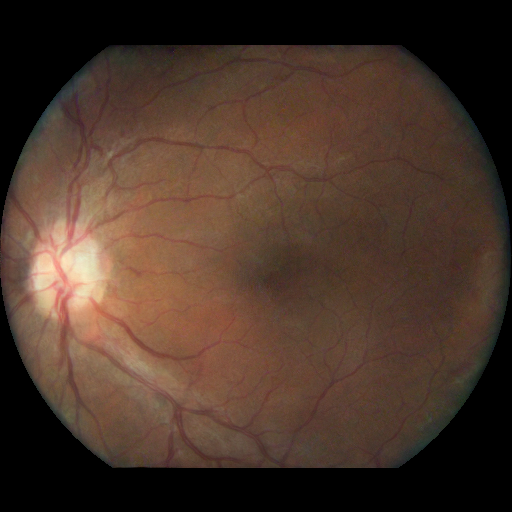}
    \includegraphics[width=0.1\textwidth]{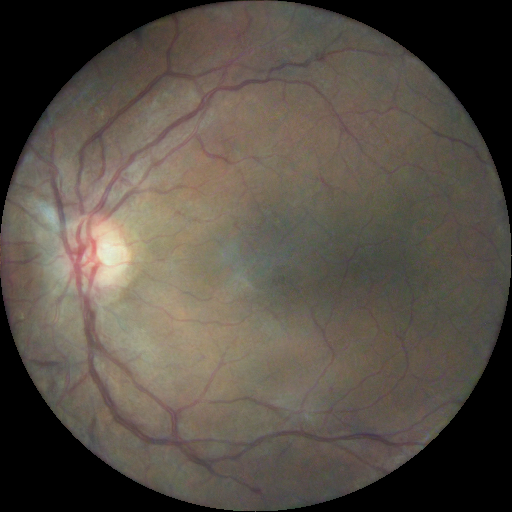}

    \vspace{0.05cm}

    \includegraphics[width=0.1\textwidth]{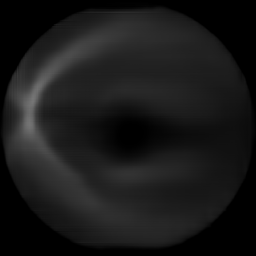}
    \includegraphics[width=0.1\textwidth]{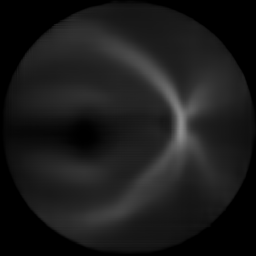}
    \hspace{0.01cm}
    \includegraphics[width=0.1\textwidth]{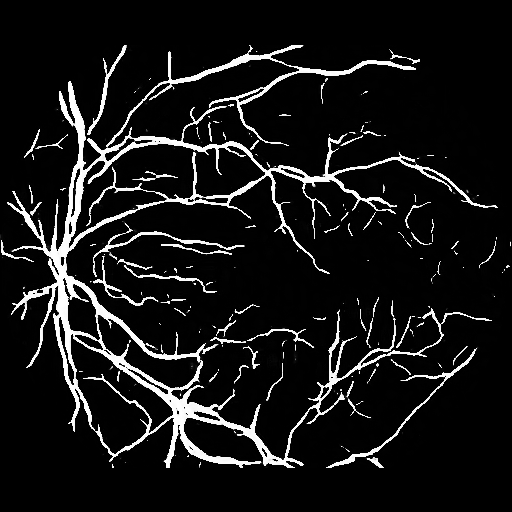}
    \includegraphics[width=0.1\textwidth]{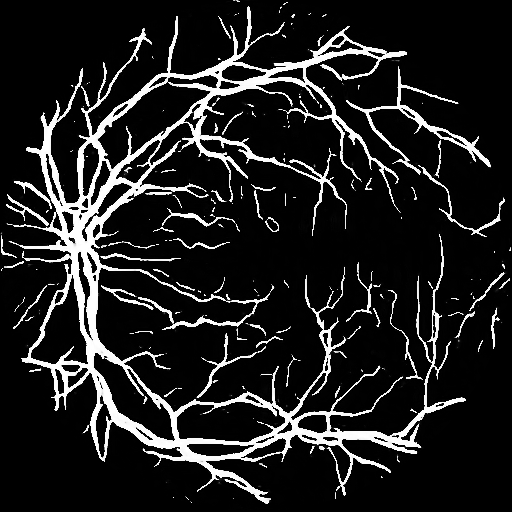}
    
    


    
    \caption{The qualitative comparison of the proposed method (on the right) with another GAN-based \cite{costa2017end} solution (on the left). The GAN could not converge, leading to low-quality results compared to ours. Alternatively, images generated by our DDPM have an accurate vessel and disk/cup structure and realistic and diverse colors.}
    \label{fig:qualitative2}
\end{figure}


\begin{figure*}[htb]
    \centering
    \begin{minipage}{1\textwidth}
    \hspace{0.95cm} Image \hspace{1.5cm} G.T. \hspace{1.7cm} O.D. \hspace{1.4cm} GAN \cite{costa2017end} \hspace{1.2cm} Ours \hspace{0.7cm} GAN \cite{costa2017end} + O.D. \hspace{0.3cm} Ours+O.D.
    \end{minipage}\vspace{0.07cm}
    \includegraphics[width=0.135\textwidth]{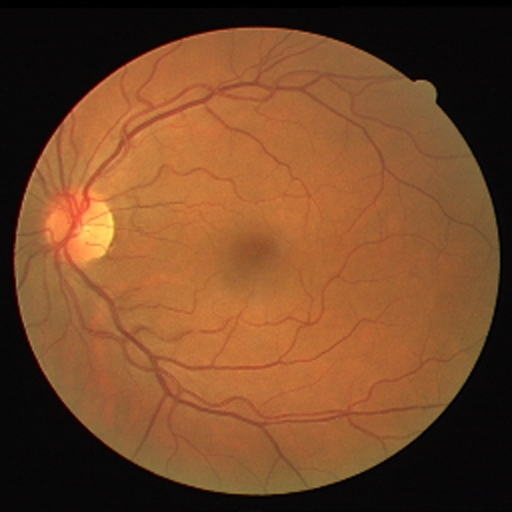}
    \includegraphics[width=0.135\textwidth]{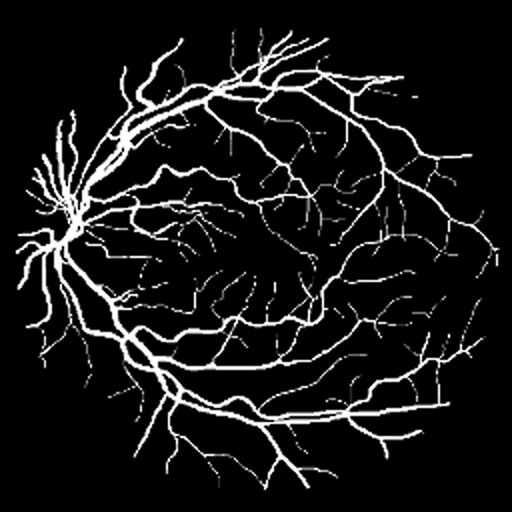}
    \includegraphics[width=0.135\textwidth]{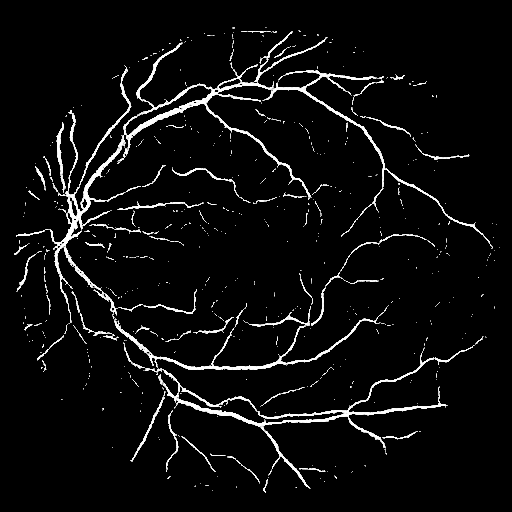}
    \includegraphics[width=0.135\textwidth]{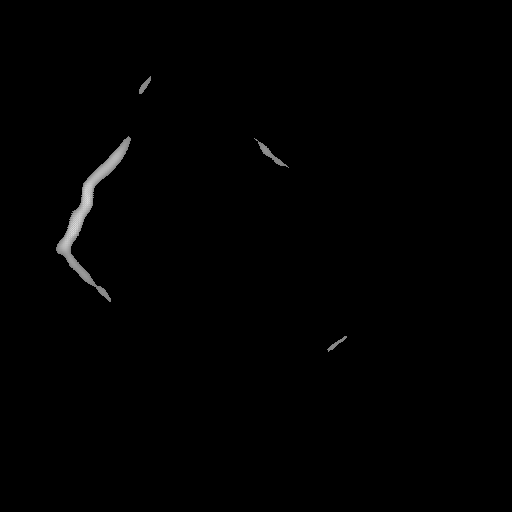}
    \includegraphics[width=0.135\textwidth]{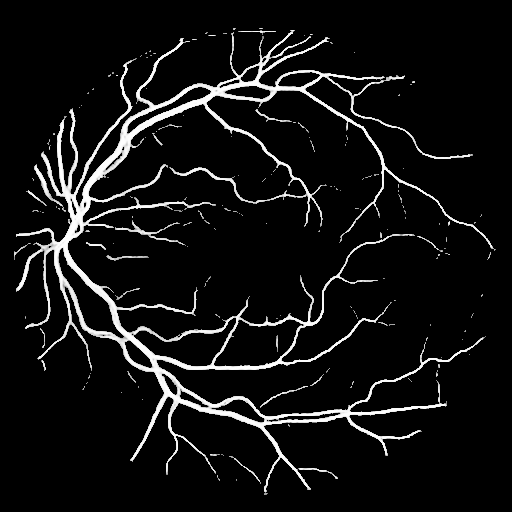}
    \includegraphics[width=0.135\textwidth]{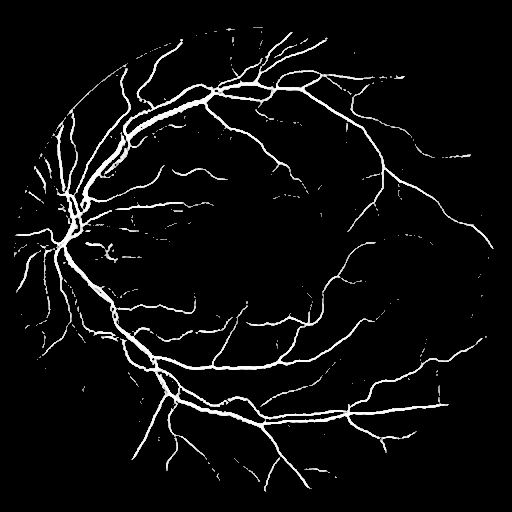}
    \includegraphics[width=0.135\textwidth]{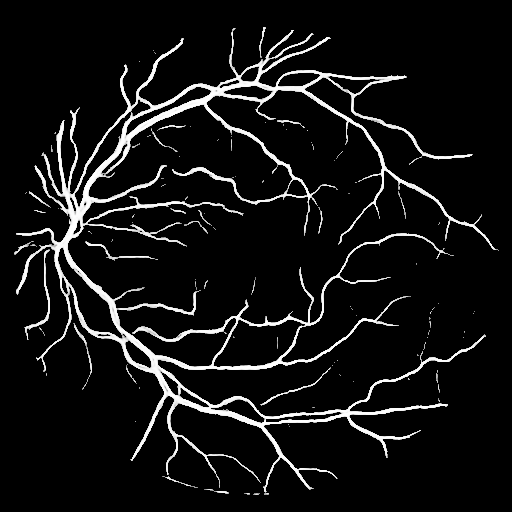}

    \vspace{0.05cm}

    \includegraphics[width=0.135\textwidth]{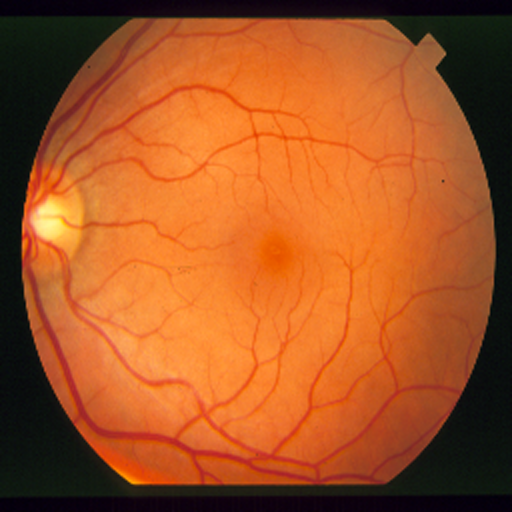}
    \includegraphics[width=0.135\textwidth]{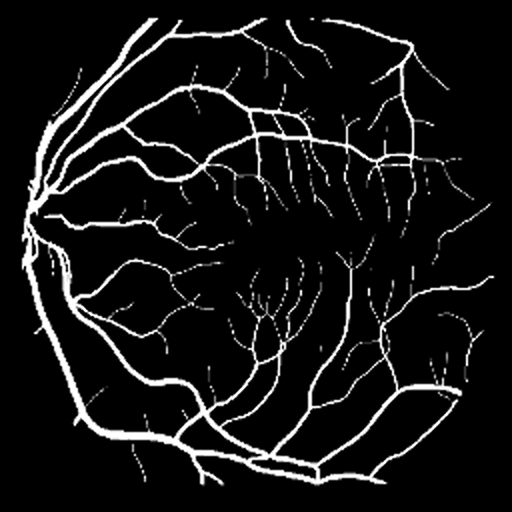}
    \includegraphics[width=0.135\textwidth]{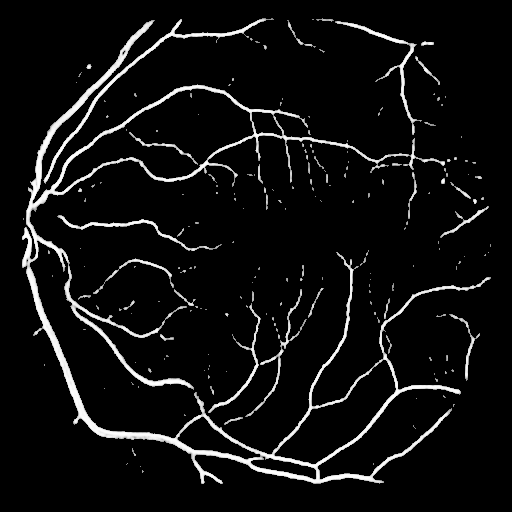}
    \includegraphics[width=0.135\textwidth]{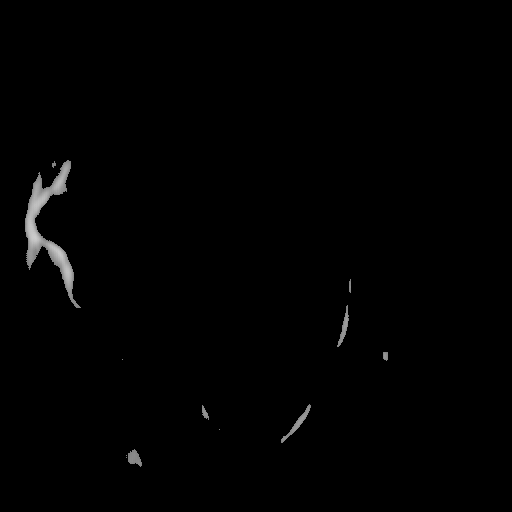}
    \includegraphics[width=0.135\textwidth]{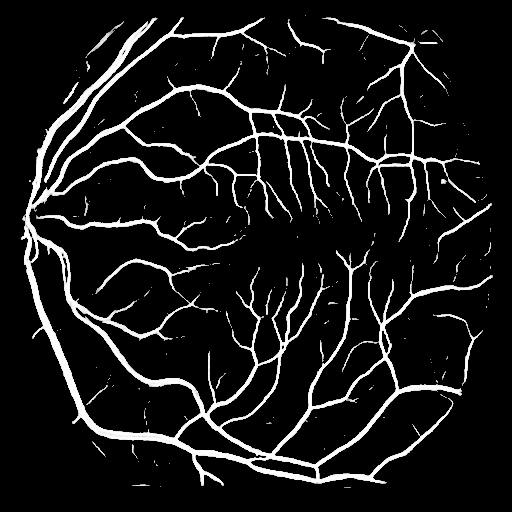}
    \includegraphics[width=0.135\textwidth]{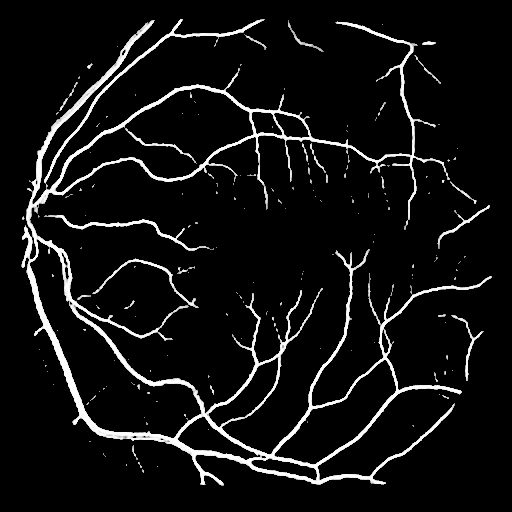}
    \includegraphics[width=0.135\textwidth]{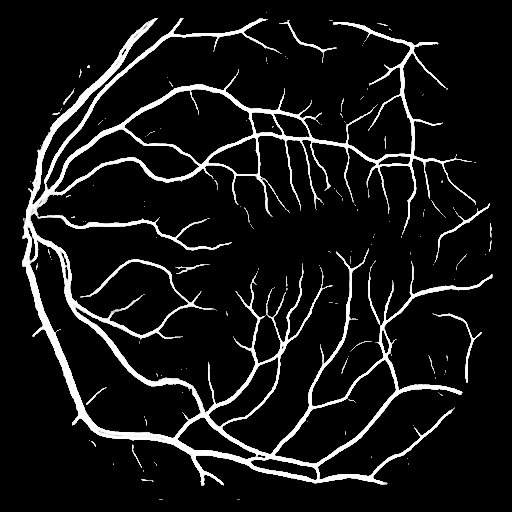}

    \caption{{The qualitative comparison of the segmentation results. From left to right: original retinal image, Ground Truth, segmentation results of UNet \cite{ronneberger2015u} trained with the original dataset, images generated by GAN \cite{costa2017end}, images generated by our method, GAN \cite{costa2017end} + original dataset combined, and ours + original dataset combined. As we can observe, our method performs best when combined with the original dataset.}}
    \label{fig:qualitative3}
\end{figure*}

\setlength{\extrarowheight}{2pt}
\begin{table}[htb]
\caption{Quantitative comparison with GAN \cite{costa2017end} in terms of Frechet Inception Distance (FID) and Single Image Generation Time (SIGT) in seconds.}
\label{table:quantitative1}
\centering
\scalebox{0.85}{
 \begin{tabular}{ |c c c| } 
 \hline
 \textbf{Method} & \textbf{FID} & \textbf{SIGT (sec)} \\ [0.5ex]
 \hline
 GAN \cite{costa2017end} & 162.50 & 6.90 \\ [0.5ex]
  \hline
 Ours & \textbf{48.45} & \textbf{6.23} \\ [0.5ex]
 \hline
 \end{tabular}}
\end{table}

\setlength{\extrarowheight}{2pt}

\begin{table*}[htb]
\caption{Quantitative comparison for vessel segmentation using DRIVE \cite{staal2004ridge}, STARE \cite{hoover2000locating}, CHASE DB1 \cite{fraz2012ensemble} datasets. {First, we combined original training sets with GAN \cite{costa2017end} and ours to train UNet \cite{ronneberger2015u}. Next, we trained UNet \cite{ronneberger2015u} with images generated by GAN \cite{costa2017end} and our methods without original data. Finally, we trained UNet \cite{ronneberger2015u} with original datasets. The results were evaluated with testing data from corresponding original datasets.} The \textbf{bold} and \underline{underlined} numbers represent the best and second-best results.}
\label{table:quantitative2}
\centering
\scalebox{0.85}{
 \begin{tabular}{ |c  c  c  c  c  c  c  c  c| } 
 \hline
 \textbf{Test data} & \textbf{Train data} & \textbf{Jaccard $\uparrow$} & \textbf{MCC $\uparrow$} & \textbf{Kappa $\uparrow$} & \textbf{F1-score $\uparrow$} & \textbf{Precision $\uparrow$} & \textbf{Recall $\uparrow$} & \textbf{Accuracy $\uparrow$} \\ [0.5ex]
 \hline
 \hline
 \multirow{6}{*}{\specialcell{\rotatebox[origin=c]{90}{DRIVE \cite{staal2004ridge}}}} 
 & DRIVE \cite{staal2004ridge} + GAN \cite{costa2017end} & 0.4419 & 0.6233 & 0.5833 & 0.6065 & \textbf{0.9017} & 0.4662 & 0.9496 \\ [0.5ex]
 & DRIVE \cite{staal2004ridge} + ReTree & \textbf{0.6161} & \textbf{0.7449} & \textbf{0.7414} & \textbf{0.7620} & \underline{0.8601} & \underline{0.7162} & \textbf{0.9723} \\ [0.5ex]
 & GAN \cite{costa2017end}  & 0.0885 & 0.2252 & 0.1430 & 0.1585 & 0.7497 & 0.0928 & 0.9190 \\ [0.5ex]
 & ReTree & \underline{0.6023} & \underline{0.7389} & \underline{0.7305} & \underline{0.7506} & 0.6714 & \textbf{0.8596} & \underline{0.9621} \\ [0.5ex]
 & DRIVE \cite{staal2004ridge} & 0.5042 & 0.6639 & 0.6445 & 0.6681 & 0.8517 & 0.5563 & 0.9532 \\ [0.5ex]
 
 \hline
 
 \multirow{6}{*}{\specialcell{\rotatebox[origin=c]{90}{STARE \cite{hoover2000locating}}}} 
 & STARE \cite{hoover2000locating} + GAN \cite{costa2017end} & \underline{0.5094} & \underline{0.6570} & \underline{0.6410} & \underline{0.6616} & \textbf{0.7912} & 0.5998 & \underline{0.9581} \\ [0.5ex]
 & STARE \cite{hoover2000locating} + ReTree & \textbf{0.5864} & \textbf{0.7240} & \textbf{0.7162} & \textbf{0.7355} & \underline{0.7900} & \textbf{0.7303} & \textbf{0.9632} \\ [0.5ex]
 & GAN \cite{costa2017end} & 0.0589 & 0.1521 & 0.0928 & 0.1084 & 0.5548 & 0.0654 & 0.9238 \\ [0.5ex]
 & ReTree & 0.4929 & 0.6324 & 0.6218 & 0.6455 & 0.7015 & \underline{0.6324} & 0.9531 \\ [0.5ex]
 & STARE \cite{hoover2000locating} & 0.4557 & 0.5918 & 0.5840 & 0.6123 & 0.6474 & 0.6068 & 0.9456 \\ [0.5ex]
 
 \hline
 \multirow{6}{*}{\specialcell{\rotatebox[origin=c]{90}{CHASE DB1 \cite{fraz2012ensemble}}}}
 & CHASE DB1 \cite{fraz2012ensemble} + GAN \cite{costa2017end} & \underline{0.4593} & \underline{0.6052} & \underline{0.6040} & \underline{0.6291} & 0.6413 & 0.6210 & \underline{0.9530} \\ [0.5ex]
 & CHASE DB1 \cite{fraz2012ensemble} + ReTree & \textbf{0.5394} & \textbf{0.6686} & \textbf{0.6655} & \textbf{0.6992} & \textbf{0.7194} & \textbf{0.7007} & \textbf{0.9649} \\ [0.5ex]
 & GAN \cite{costa2017end} & 0.0913 & 0.2277 & 0.1519 & 0.1663 & 0.6320 & 0.0972 & 0.9380 \\ [0.5ex]
 & ReTree & 0.3580 & 0.5146 & 0.5005 & 0.5256 & \underline{0.6700} & 0.4355 & 0.9497 \\ [0.5ex]
 & CHASE DB1 \cite{fraz2012ensemble} & 0.4319 & 0.5793 & 0.5716 & 0.6030 & 0.5308 & \underline{0.7033} & 0.9405 \\ [0.5ex]
 \hline
 \end{tabular}}
\end{table*}

\subsection{Qualitative Results}

In our experiments, we compared our method with GAN by Costa et al. \cite{costa2017end}, {since the authors of most of the other methods that focused on retinal image generation did not publish their implementations and generated datasets. As a result, we could not compare our method with these works as we did not receive any positive reply from the corresponding authors. However, we compared the proposed method with Improved DDPM (IDDPM) \cite{nichol2021improved} in the ablation study.} In addition, the proposed model does not suffer from the mode collapse that is present in GANs; as a result, the images obtained by our model are diverse in terms of color and overall structure. Fig. \ref{fig:qualitative2} shows the qualitative comparison with another GAN-based method \cite{costa2017end}; as we can observe, the GAN-based model has failed to converge since the generated images lack clear visual vessel structure. In addition, the generated vessel maps need more visual clarity to be used for retinal segmentation tasks.

Additionally, we validated the proposed ReTree dataset using UNet \cite{ronneberger2015u}. To do so, we trained UNet with generated images from the ReTree dataset. Then we tested it using real images from three publicly available datasets such as DRIVE \cite{staal2004ridge}, STARE \cite{hoover2000locating}, and CHASE DB1 \cite{fraz2012ensemble}. The qualitative results are presented in Fig. \ref{fig:qualitative3}. As we can observe, using ReTree images and images from original datasets is best, leading to better qualitative results. As we can also observe, the results of GAN method \cite{costa2017end} failed in segmentation, while GAN \cite{costa2017end} + original dataset could improve the segmentation process. When training with the ReTree dataset, UNet still performs well in vessel segmentation; however, it is outperformed by ReTree + original dataset.

\setlength{\extrarowheight}{2pt}
\begin{table}[htb]
\caption{Quantitative comparison with original IDDPM \cite{nichol2021improved}, IDDPM + RTT, ours without RTT and ours + RTT in terms of Frechet Inception Distance (FID), Single Image Generation Time (SIGT) and training time (in hours) for two stages of generation.}
\label{table:ablation1}
\centering
\scalebox{0.85}{
 \begin{tabular}{ |c c c c| } 
 \hline
 \textbf{Method} & \textbf{FID} & \textbf{SIGT (sec)} & \textbf{Train time (hours)} \\ [0.5ex]
 \hline
 \hline
 IDDPM \cite{nichol2021improved} w\textbackslash{o} RTT & 68.959 & 6.156 + 6.375 & 8.8 + 7.4 \\ [0.5ex]
 \hline
 IDDPM \cite{nichol2021improved} w RTT & 64.32 & 6.156 + 6.375 & 5.4 + 5.1 \\ [0.5ex]
 \hline
 Ours w\textbackslash{o} RTT & 55.49 & 2.97 + 3.26 & 7.07 + 7.2 \\ [0.5ex]
 \hline
 Ours w RTT & \textbf{48.45} & \textbf{2.97 + 3.26} & \textbf{4.29 + 4.01} \\ [0.5ex]
 \hline
 \end{tabular}}
\end{table}

\setlength{\extrarowheight}{2pt}

\begin{table*}[htb]
\caption{Quantitative results comparison for vessel segmentation using DRIVE \cite{staal2004ridge}, STARE \cite{hoover2000locating}, CHASE DB1 \cite{fraz2012ensemble} datasets. {First, we trained the super-resolution network to up-scale generated images by factors of $\times8$ and $\times4$ (col. 1). Then, we combined original training datasets (col. 2) with images generated by models from the ablation study (col. 3) to train UNet \cite{ronneberger2015u}. Next, we evaluated the results using test sets of corresponding original datasets.} The \textbf{bold} and \underline{underlined} numbers represent the best and second-best results.}
\label{table:ablation2}
\centering
\scalebox{0.7}{
 \begin{tabular}{ |c | c  c  c  c  c  c  c  c  c| } 
 \hline
 \textbf{Resolution} & \textbf{Test Data} & \textbf{Train data} & \textbf{Jaccard $\uparrow$} & \textbf{MCC $\uparrow$} & \textbf{Kappa $\uparrow$} & \textbf{F1-score $\uparrow$} & \textbf{Precision $\uparrow$} & \textbf{Recall $\uparrow$} & \textbf{Accuracy $\uparrow$} \\ [0.5ex]
 \hline

 \multirow{28}{*}{\specialcell{\rotatebox[origin=c]{90}{512$\times$512}}}
 & \multirow{10}{*}{\specialcell{\rotatebox[origin=c]{90}{DRIVE \cite{staal2004ridge}}}} 
 & DRIVE \cite{staal2004ridge} + IDDPM \cite{nichol2021improved} w\textbackslash{o} RTT & \underline{0.6103} & \underline{0.7417} & \underline{0.7369} & \underline{0.7574} & 0.8335 & 0.6981 & \underline{0.9618}\\ [0.5ex]
 & & DRIVE \cite{staal2004ridge} + IDDPM \cite{nichol2021improved} w RTT & 0.5605 & 0.7077 & 0.6956 & 0.7174 & \underline{0.8575} & 0.6207 & 0.9582 \\ [0.5ex]
 & & {IDDPM \cite{nichol2021improved} w\textbackslash{o} RTT} & 0.4703 & 0.6073 & 0.6054 & 0.6390 & 0.6423 & 0.6427 & 0.9381 \\ [0.5ex]
 & & {IDDPM \cite{nichol2021improved} w RTT} & 0.4756 & 0.6128 & 0.6109 & 0.6437 & 0.6544 & 0.6395 & 0.9397 \\ [0.5ex]
 & & DRIVE \cite{staal2004ridge} + Ours w\textbackslash{o} RTT & 0.5874 & 0.7266 & 0.7185 & 0.7395 & 0.8505 & 0.6573 & 0.9604 \\ [0.5ex]
 & & DRIVE \cite{staal2004ridge} + Ours w RTT & \textbf{0.6161} & \textbf{0.7449} & \textbf{0.7414} & \textbf{0.7620} & \textbf{0.8601} & \textbf{0.7162} & \textbf{0.9723} \\ [0.5ex]
 & & {Ours w\textbackslash{o} RTT} & 0.5046 & 0.6402 & 0.6373 & 0.6700 & 0.6358 & \underline{0.7145} & 0.9399 \\ [0.5ex]
 & & {Ours w RTT} & 0.5208 & 0.6565 & 0.6549 & 0.6841 & 0.6917 & 0.6824 & 0.9463 \\ [0.5ex]
 
 \cline{2-10}
 
 & \multirow{10}{*}{\specialcell{\rotatebox[origin=c]{90}{STARE \cite{hoover2000locating}}}} 
 & STARE \cite{hoover2000locating} + IDDPM \cite{nichol2021improved} w\textbackslash{o} RTT & \underline{0.5860} & \underline{0.7226} & \underline{0.7152} & \underline{0.7350} & \underline{0.7712} & 0.7272 & \underline{0.9623} \\ [0.5ex]
 & & STARE \cite{hoover2000locating} + IDDPM \cite{nichol2021improved} w RTT & 0.5574 & 0.6938 & 0.6894 & 0.7122 & 0.7201 & 0.7208 & 0.9573 \\ [0.5ex]
 & & {IDDPM \cite{nichol2021improved} w\textbackslash{o} RTT} & 0.2964 & 0.4215 & 0.4059 & 0.4376 & 0.5510 & 0.4237 & 0.9297 \\ [0.5ex]
 & & {IDDPM \cite{nichol2021improved} w RTT} & 0.4716 & 0.6171 & 0.6079 & 0.6395 & 0.5974 & 0.7145 & 0.9404 \\ [0.5ex]
 & & STARE \cite{hoover2000locating} + Ours w\textbackslash{o} RTT & 0.5767 & 0.7128 & 0.7082 & 0.7296 & 0.7473 & \underline{0.7284} & 0.9600 \\ [0.5ex]
 & & STARE \cite{hoover2000locating} + Ours w RTT & \textbf{0.5864} & \textbf{0.7240} & \textbf{0.7162} & \textbf{0.7355} & \textbf{0.7900} & \textbf{0.7303} & \textbf{0.9632} \\ [0.5ex]
 & & {Ours w\textbackslash{o} RTT} & 0.3910 & 0.5327 & 0.5195 & 0.5546 & 0.5598 & 0.5998 & 0.9318 \\ [0.5ex]
 & & {Ours w RTT} & 0.4970 & 0.6403 & 0.6339 & 0.6630 & 0.6240 & 0.7247 & 0.9456 \\ [0.5ex]
 
 \cline{2-10}

 & \multirow{10}{*}{\specialcell{\rotatebox[origin=c]{90}{CHASE \cite{fraz2012ensemble}}}}
 & CHASE \cite{fraz2012ensemble} + IDDPM \cite{nichol2021improved} w\textbackslash{o} RTT & 0.4709 & 0.6196 & 0.6169 & 0.6397 & 0.6909 & 0.5985 & 0.9567 \\ [0.5ex]
 & & CHASE \cite{fraz2012ensemble} + IDDPM \cite{nichol2021improved} w RTT & \underline{0.5161} & \underline{0.6590} & \underline{0.6577} & \underline{0.6806} & \underline{0.7093} & 0.6566 & \underline{0.9572} \\ [0.5ex]
 & & {IDDPM \cite{nichol2021improved} w\textbackslash{o} RTT} & 0.4201 & 0.5643 & 0.5559 & 0.5883 & 0.5473 & 0.6830 & 0.9401 \\ [0.5ex]
 & & {IDDPM \cite{nichol2021improved} w RTT} & 0.4484 & 0.5947 & 0.5898 & 0.6190 & 0.5597 & \underline{0.6963} & 0.9449 \\ [0.5ex]
 & & CHASE \cite{fraz2012ensemble} + Ours w\textbackslash{o} RTT & 0.4527 & 0.6003 & 0.5987 & 0.6230 & 0.6640 & 0.5880 & 0.9542 \\ [0.5ex]
 & & CHASE \cite{fraz2012ensemble} + Ours w RTT & \textbf{0.5394} & \textbf{0.6686} & \textbf{0.6655} & \textbf{0.6992} & \textbf{0.7194} & \textbf{0.7007} & \textbf{0.9649} \\ [0.5ex]
 & & {Ours w\textbackslash{o} RTT} & 0.4218 & 0.5670 & 0.5616 & 0.5931 & 0.5328 & 0.6730 & 0.9406 \\ [0.5ex]
 & & {Ours w RTT} & 0.4625 & 0.6074 & 0.6060 & 0.6322 & 0.6124 & 0.6570 & 0.9509 \\ [0.5ex]
 
 \hline

 \multirow{28}{*}{\specialcell{\rotatebox[origin=c]{90}{256$\times$256}}}
 & \multirow{10}{*}{\specialcell{\rotatebox[origin=c]{90}{DRIVE \cite{staal2004ridge}}}} 
 & DRIVE \cite{staal2004ridge} + IDDPM \cite{nichol2021improved} w\textbackslash{o} RTT & 0.6334 & 0.7572 & 0.7551 & 0.7754 & \textbf{0.8151} & 0.7429 & 0.9626 \\ [0.5ex]
 & & DRIVE \cite{staal2004ridge} + IDDPM \cite{nichol2021improved} w RTT & \underline{0.6472} & \underline{0.7649} & \underline{0.7637} & \underline{0.7830} & \underline{0.8081} & \underline{0.7723} & \textbf{0.9635} \\ [0.5ex]
 & & {IDDPM \cite{nichol2021improved} w\textbackslash{o} RTT} & 0.5143 & 0.6531 & 0.6521 & 0.6813 & 0.6941 & 0.6917 & 0.9497 \\ [0.5ex]
 & & {IDDPM \cite{nichol2021improved} w RTT} & 0.5416 & 0.6691 & 0.6639 & 0.6995 & 0.7034 & 0.6855 & 0.9599 \\ [0.5ex]
 & & DRIVE \cite{staal2004ridge} + Ours w\textbackslash{o} RTT & 0.6401 & 0.7613 & 0.7598 & 0.7803 & 0.7961 & 0.7695 & 0.9624 \\ [0.5ex]
 & & DRIVE \cite{staal2004ridge} + Ours w RTT & \textbf{0.6486} & \textbf{0.7684} & \textbf{0.7667} & \textbf{0.7866} & 0.8000 & \textbf{0.7794} & \underline{0.9633} \\ [0.5ex]
 & & {Ours w\textbackslash{o} RTT} & 0.5543 & 0.7062 & 0.6993 & 0.7271 & 0.6993 & 0.7528 & 0.9594 \\ [0.5ex]
 & & {Ours w RTT} & 0.5601 & 0.7085 & 0.7041 & 0.7342 & 0.7387 & 0.7464 & 0.9593 \\ [0.5ex]
 
 \cline{2-10}
 
 & \multirow{10}{*}{\specialcell{\rotatebox[origin=c]{90}{STARE \cite{hoover2000locating}}}} 
 & STARE + IDDPM \cite{nichol2021improved} w\textbackslash{o} RTT & 0.5780 & 0.7105 & 0.7080 & 0.7282 & \underline{0.7618} & 0.7037 & 0.9610 \\ [0.5ex]
 & & STARE + IDDPM \cite{nichol2021improved} w RTT & 0.5844 & \underline{0.7196} & \underline{0.7156} & \underline{0.7355} & \textbf{0.7900} & 0.6957 & \underline{0.9627} \\ [0.5ex]
 & & {IDDPM \cite{nichol2021improved} w\textbackslash{o} RTT} & 0.4124 & 0.4511 & 0.4383 & 0.4672 & 0.5881 & 0.4617 & 0.9537 \\ [0.5ex]
 & & {IDDPM \cite{nichol2021improved} w RTT} & 0.5156 & 0.6421 & 0.6409 & 0.6691 & 0.6311 & \underline{0.7415} & 0.9594 \\ [0.5ex]
 & & STARE + Ours w\textbackslash{o} RTT & \underline{0.5855} & 0.7161 & 0.7143 & 0.7347 & 0.7434 & 0.7325 & 0.9621 \\ [0.5ex]
 & & STARE + Ours w RTT & \textbf{0.5869} & \textbf{0.7201} & \textbf{0.7170} & \textbf{0.7384} & 0.7558 & 0.7322 & \textbf{0.9671} \\ [0.5ex]
 & & {Ours w\textbackslash{o} RTT} & 0.4381 & 0.5687 & 0.5415 & 0.5811 & 0.5871 & 0.6251 & 0.9578 \\ [0.5ex]
 & & {Ours w RTT} & 0.5411 & 0.6716 & 0.6719 & 0.6916 & 0.6571 & \textbf{0.7527} & 0.9596 \\ [0.5ex]
 
 \cline{2-10}

 & \multirow{10}{*}{\specialcell{\rotatebox[origin=c]{90}{CHASE \cite{fraz2012ensemble}}}}
 & CHASE \cite{fraz2012ensemble} + IDDPM \cite{nichol2021improved} w\textbackslash{o} RTT & 0.5570 & 0.6962 & 0.6950 & 0.7153 & 0.6948 & 0.7400 & 0.9621 \\ [0.5ex]
 & & CHASE \cite{fraz2012ensemble} + IDDPM \cite{nichol2021improved} w RTT & \underline{0.5855} & \underline{0.7218} & \underline{0.7191} & \textbf{0.7384} & 0.6929 & \textbf{0.7934} & \underline{0.9639} \\ [0.5ex]
 & & {IDDPM \cite{nichol2021improved} w\textbackslash{o} RTT} & 0.4810 & 0.6128 & 0.6101 & 0.6431 & 0.6008 & 0.7451 & 0.9581 \\ [0.5ex]
 & & {IDDPM \cite{nichol2021improved} w RTT} & 0.5095 & 0.6511 & 0.6481 & 0.6731 & 0.6251 & 0.7531 & 0.9601 \\ [0.5ex]
 & & CHASE \cite{fraz2012ensemble} + Ours w\textbackslash{o} RTT & 0.5697 & 0.7076 & 0.7061 & 0.7257 & \underline{0.6999} & 0.7567 & 0.9633 \\ [0.5ex]
 & & CHASE \cite{fraz2012ensemble} + Ours w RTT & \textbf{0.5889} & \textbf{0.7267} & \textbf{0.7254} & \underline{0.7351} & \textbf{0.7088} & \underline{0.7661} & \textbf{0.9642} \\ [0.5ex]
 & & {Ours w\textbackslash{o} RTT} & 0.4891 & 0.6231 & 0.6216 & 0.6507 & 0.5931 & 0.7328 & 0.9606 \\ [0.5ex]
 & & {Ours w RTT} & 0.5310 & 0.6631 & 0.6647 & 0.6942 & 0.6785 & 0.7231 & 0.9623 \\ [0.5ex]
 
 \hline
 \end{tabular}}
\end{table*}

\begin{figure*}[htb]
    \centering
    \begin{minipage}{1\textwidth}
    \hspace{1.9cm} Image \hspace{1.7cm} G.T. \hspace{1.2cm} IDDPM \cite{nichol2021improved} \hspace{0.25cm} IDDPM \cite{nichol2021improved} + RTT \hspace{0.65cm} Ours \hspace{1.2cm} Ours + RTT
    \end{minipage}\vspace{0.07cm}
    \includegraphics[width=0.14\textwidth]{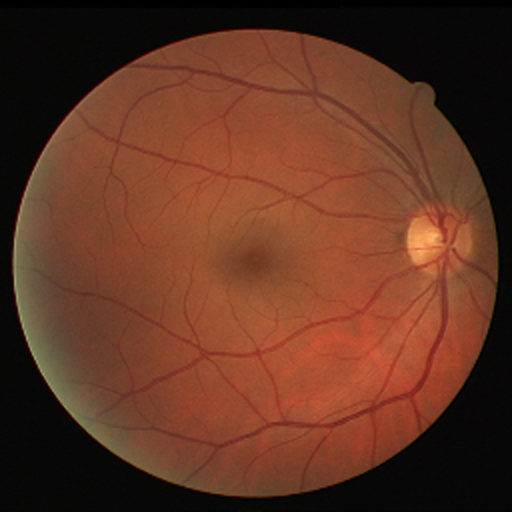}
    \includegraphics[width=0.14\textwidth]{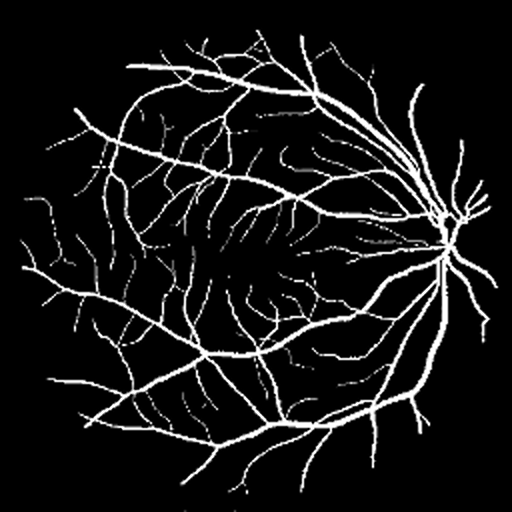}
    \includegraphics[width=0.14\textwidth]{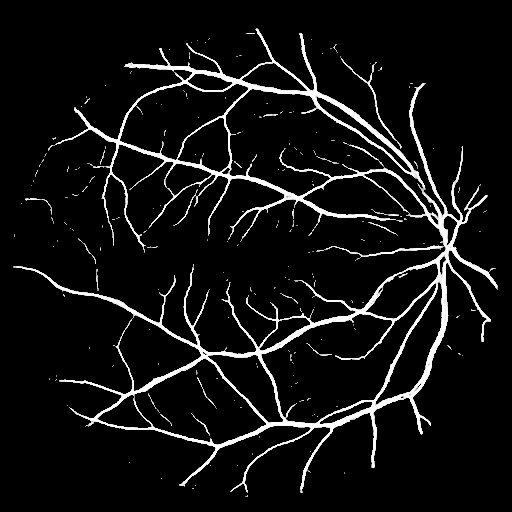}
    \includegraphics[width=0.14\textwidth]{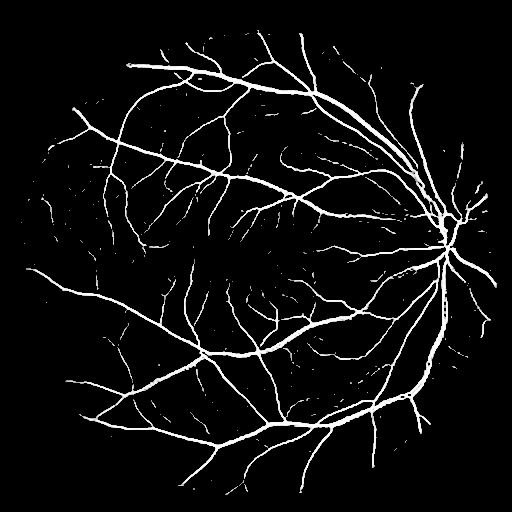}
    \includegraphics[width=0.14\textwidth]{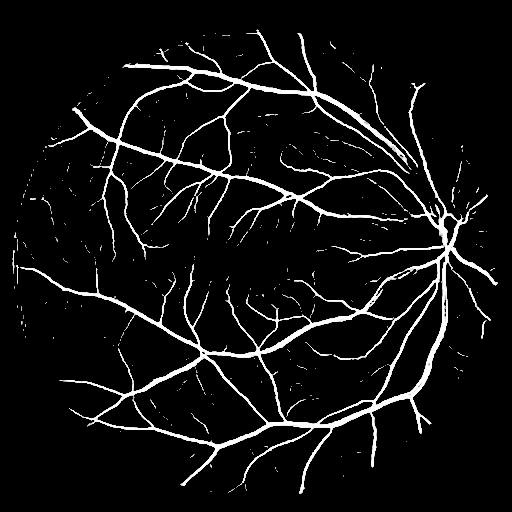}
    \includegraphics[width=0.14\textwidth]{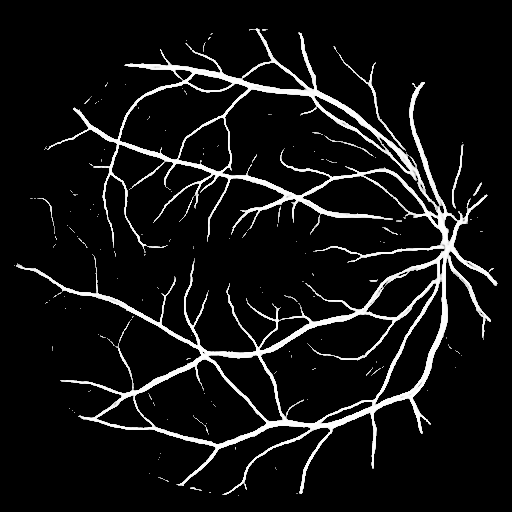}

    \vspace{0.05cm}

    \includegraphics[width=0.14\textwidth]{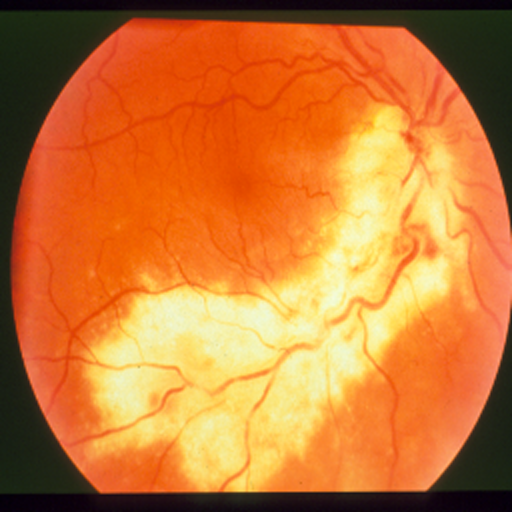}
    \includegraphics[width=0.14\textwidth]{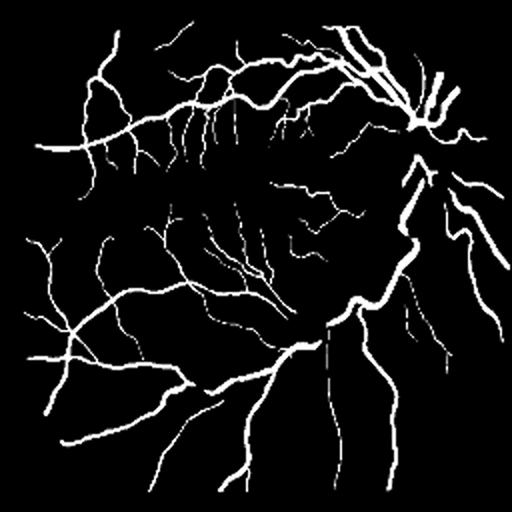}
    \includegraphics[width=0.14\textwidth]{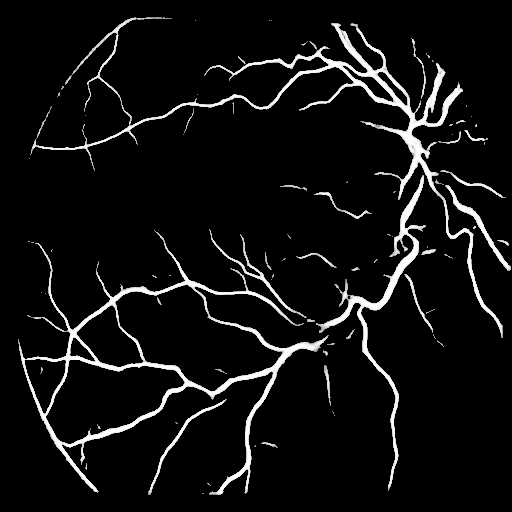}
    \includegraphics[width=0.14\textwidth]{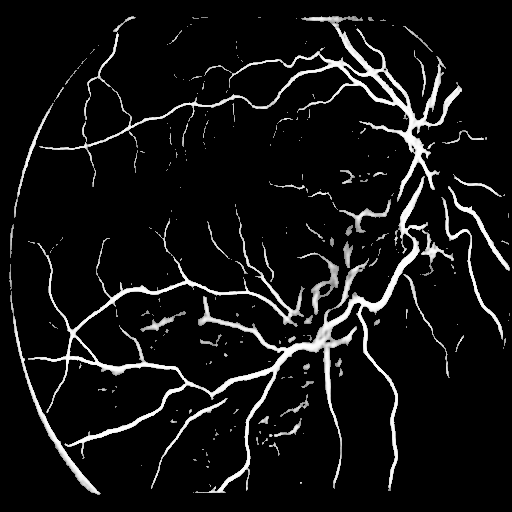}
    \includegraphics[width=0.14\textwidth]{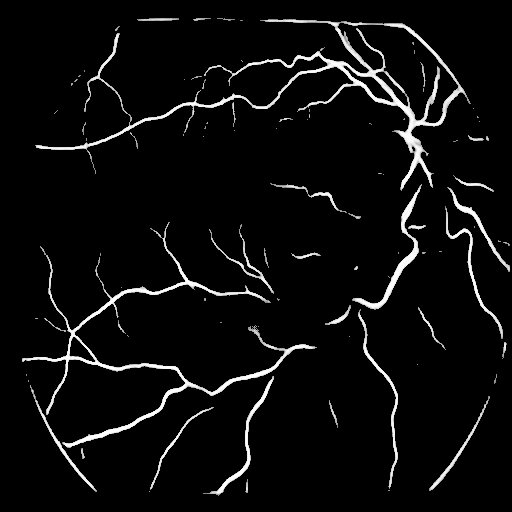}
    \includegraphics[width=0.14\textwidth]{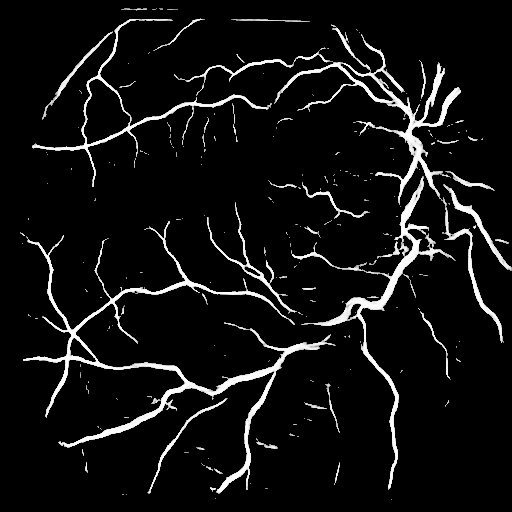}
    
    \caption{{The qualitative comparison of the segmentation results. Left to right: original retinal image, Ground Truth (G.T.), segmentation results of UNet \cite{ronneberger2015u} trained with images generated by IDDPM \cite{nichol2021improved}, IDDPM \cite{nichol2021improved} + original dataset combined, ours, and ours + original dataset combined. As we can observe, our method performs best when combined with the original dataset.}}
    \label{fig:ablation}
\end{figure*}

\subsection{Quantitative Results}
Table \ref{table:quantitative1} shows the FID scores along with single image generation time (SIGT) obtained by GAN-based \cite{costa2017end} and our models. As we can observe, the proposed method outperforms the other techniques in having lower FID; in particular, the FI distance between ReTree and EyeQ images is more than 3 times less than that of the GAN-based method. In addition, GAN takes 6.90 seconds to generate a single image pair, while ours generates a single pair in 6.23 seconds. The proposed architecture significantly decreased the generation time by a factor of 2 compared to IDDPM \cite{nichol2021improved}.

Table \ref{table:quantitative2} represents the second quantitative evaluation stage for the proposed ReTree dataset. In this stage, we performed retinal vessel segmentation using UNet \cite{ronneberger2015u}. First, it was trained with original images from three publicly available datasets, such as DRIVE \cite{staal2004ridge}, STARE \cite{hoover2000locating}, and CHASE DB1 \cite{fraz2012ensemble}. Next, it was trained with generated images only and then generated + original images from each dataset. The three UNet versions are tested with original images from three datasets during the evaluation. As we can observe from Table \ref{table:quantitative2}, the proposed dataset performs best when combined with original datasets.



\subsection{Ablation Study}
To validate the efficiency and performance of the proposed DDPM, we performed an extensive ablation study in which we compared our method with improved DDPM \cite{nichol2021improved} that was trained in the original setup {using the same two-stage generation}; next, we trained this model with the proposed Repetitive Training Technique (RTT). Additionally, we trained the proposed DDPM without RTT. The training of {UNet \cite{ronneberger2015u}} has been performed using generated {up-scaled} images with resolutions of $256\times$256 and $512\times512$ pixels {alone and combined with training data from original corresponding retinal segmentation datasets}. Table \ref{table:ablation1} shows the FID values, SIGT, and total training time. As we can see, embedding RTT in the training process significantly reduces the total training time; for instance, RTT reduces the training time by 32.7\% from 16.2 to 10.9 hours for IDDPM \cite{nichol2021improved} and by 41.8\% from 14.27 to 8.3 hours for the proposed model. Additionally, the proposed model takes less time to train and to generate a single sample; IDDPM requires 12.53 seconds which is two times higher than the proposed model. RTT also reduced the FID values for both models; however, the proposed model has the lowest FID score. 

The models have also been trained and tested using segmentation datasets. Table \ref{table:ablation2} compares segmentation results using three datasets. {We performed $\times4$ and $\times8$ super-resolution of generated images.} The same models have been trained in combination with three segmentation datasets. As we can observe, the proposed model and training technique significantly outperforms the other methods {in both super-resolution factors}. In particular, the proposed model with RTT obtained the highest values of Jaccard similarity score, MCC, Cohen's kappa, F1-score, precision, recall, and accuracy, which means our method results in the highest similarity and positive correlation between segmentation classes. IDDPM \cite{nichol2021improved} without RTT obtained second-best quantitative results in DRIVE \cite{staal2004ridge} and STARE \cite{hoover2000locating}. However, it is outperformed by the same model with RTT in CHASE DB1 \cite{fraz2012ensemble}. Fig. \ref{fig:ablation} shows the qualitative comparison between these methods; as we can observe, utilizing our architecture along with RTT leads to better qualitative results than others. Therefore, these results present the effectiveness of the proposed DDPM architecture and RTT, which lead to higher quantitative, qualitative, and computational results.

{Furthermore, we conducted experiments with RTT to examine its impact on the training of other deep learning methods. Specifically, we applied RTT to train UNet \cite{ronneberger2015u} for retinal image segmentation. The results are presented in Table \ref{table:rtt}. As observed, RTT has a positive effect on UNet, leading to an increase in almost all quantitative metrics. The most notable improvement can be seen in the STARE dataset \cite{hoover2000locating}, where all quantitative metrics show an average increase of 3.42\% while it reduces training time by an average of 28.47\% across all datasets.}

\setlength{\extrarowheight}{2pt}

\begin{table}[htb]
\caption{Quantitative evaluation of UNet \cite{ronneberger2015u} with and without Repetitive Training Technique and total training time in minutes.}
\label{table:rtt}
\centering
\scalebox{0.6}{
 \begin{tabular}{ |c  c  c  c  c  c  c  c  c  c| } 
 \hline
 \textbf{Dataset} & \textbf{RTT} & \textbf{Jaccard} & \textbf{MCC} &\textbf{Kappa} & \textbf{F1} & \textbf{Recall} & \textbf{Precision} & \textbf{Accuracy} & \textbf{Time (mins)} \\ [0.5ex]
 \hline
 \multirow{2}{*}{\specialcell{DRIVE \cite{staal2004ridge}}} 
 & \xmark & 0.6393 & 0.7624 & \textbf{0.7584} & \textbf{0.7793} & \textbf{0.7991} & 0.7733 & 0.9615 & 17.73 \\ [0.5ex]
 & \cmark & \textbf{0.6417} & \textbf{0.7640} & 0.7575 & 0.7771 & 0.7860 & \textbf{0.7945} & \textbf{0.9628} & \textbf{13.48} \\ [0.5ex] 
 \hline
 \multirow{2}{*}{\specialcell{STARE \cite{hoover2000locating}}} 
 & \xmark & 0.5712 & 0.7142 & 0.6994 & 0.7212 & 0.7524 & 0.7446 & 0.9579 & 17.98\\ [0.5ex]
 & \cmark & \textbf{0.6042} & \textbf{0.7412} & \textbf{0.7276} & \textbf{0.7464} & \textbf{0.7549} & \textbf{0.7890} & \textbf{0.9635} & \textbf{12.27} \\ [0.5ex] 
 \hline
 \multirow{2}{*}{\specialcell{CHASE DB1 \cite{fraz2012ensemble}}} 
 & \xmark & 0.5980 & 0.7320 & 0.7299 & 0.7482 & 0.7944 & \textbf{0.7098} & 0.9656 & 14.73 \\ [0.5ex]
 & \cmark & \textbf{0.6102} & \textbf{0.7434} & \textbf{0.7396} & \textbf{0.7577} & \textbf{0.8270} & 0.7018 & \textbf{0.9660} & \textbf{10.36} \\ [0.5ex] 
 \hline
 \end{tabular}}
\end{table}

\subsection{Limitations}
\label{Limitations}
\begin{figure}[htb]
    \centering
    \includegraphics[width=0.11\textwidth]{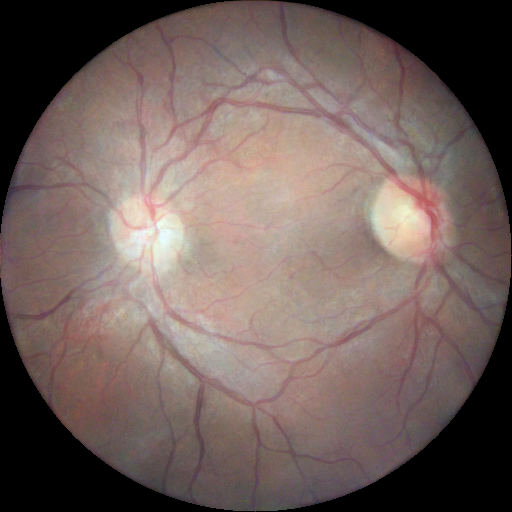}
    \includegraphics[width=0.11\textwidth]{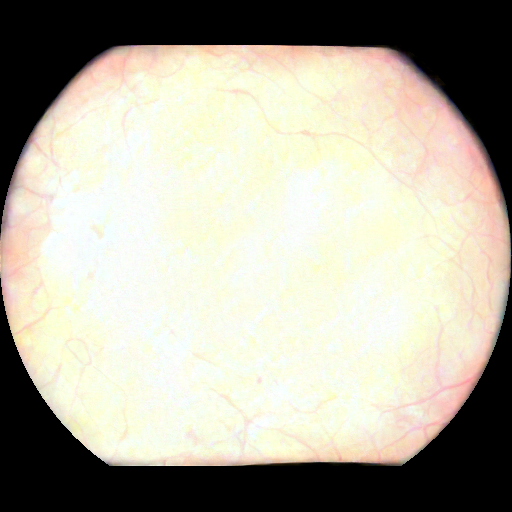}
    \includegraphics[width=0.11\textwidth]{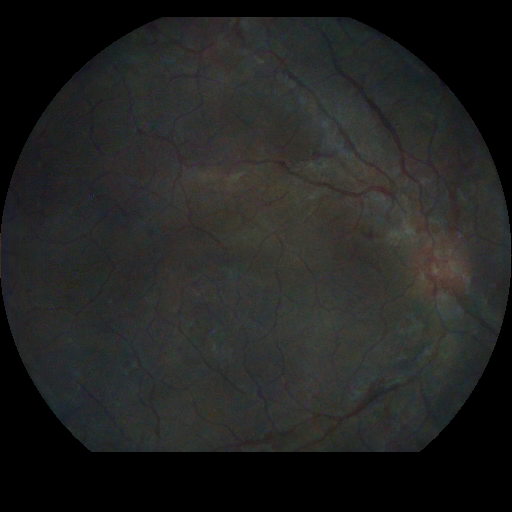}
    \includegraphics[width=0.11\textwidth]{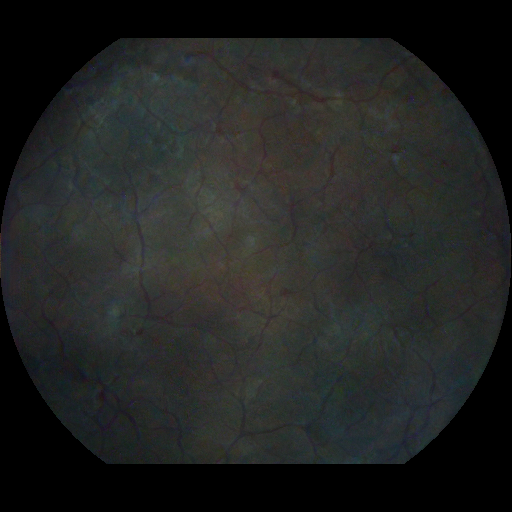}

    \vspace{0.05cm}

    \includegraphics[width=0.11\textwidth]{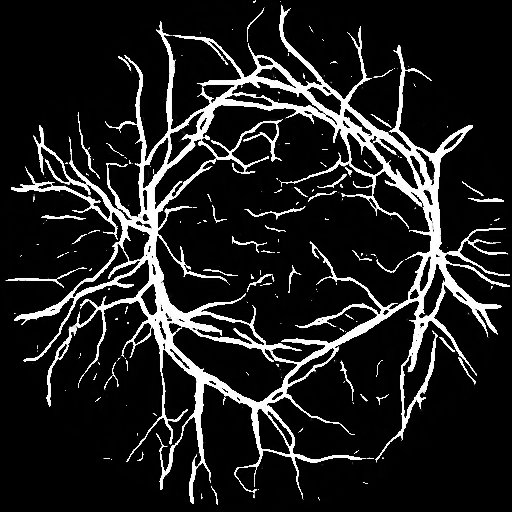}
    \includegraphics[width=0.11\textwidth]{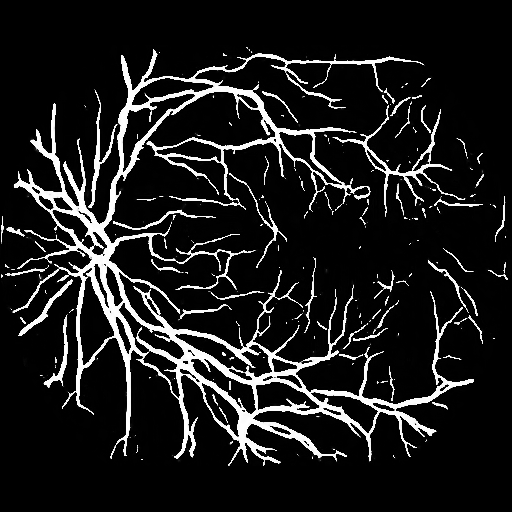}
    \includegraphics[width=0.11\textwidth]{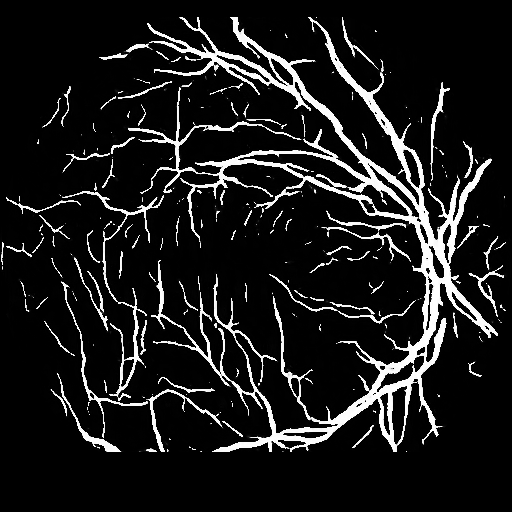}
    \includegraphics[width=0.11\textwidth]{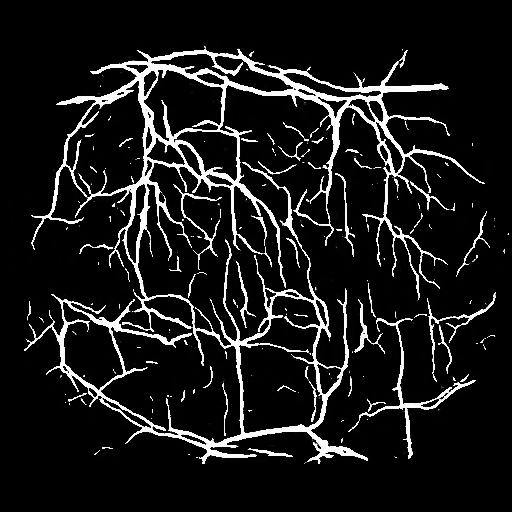}

    \vspace{0.05cm}



    
    \caption{Examples of failure cases of our proposed method including retinal images having two retinal cups, over-bright regions, dark results, and images with no retinal cup.}
    \label{fig:limitations}
\end{figure}

The proposed DDPMs can produce realistic retinal images along with vessel maps. However, since DDPMs use random noise to generate samples, they may generate unrealistic images, which is caused by the fact that DDPMs have the potential to simplify the assumptions made about the probability distribution underlying the input data, which could involve assuming a Gaussian distribution or a specific degree of smoothness. {However, these assumptions may not necessarily hold in practical scenarios, and this can lead to the generation of unrealistic images}. As shown in Fig. \ref{fig:limitations}, some images may have two retinal cups, uneven illumination, color distortion, and missing retinal cups. However, we trained a discriminative model {using real images from EyeQ \cite{fu2019evaluation} dataset and a subset of 10,000 generated images from the proposed dataset} to classify real and generated images to overcome these limitations. The output of this discriminator is a continuous prediction value between 0 and 1, where 0 is generated and 1 is a real image. The threshold has been set to 0.8, which allowed us to keep only realistically-looking retinal images in the dataset. It helped to remove the unrealistic images from the dataset.

\section{Conclusion}
\label{Conclusion}
In this work, our main objective was to develop a novel retinal image segmentation dataset generated with the current state-of-the-art class of generation models, namely Denoising Diffusion Probabilistic Models. Additionally, we propose a novel lightweight architecture and a training technique for DDPM that significantly improves the computational, qualitative, and quantitative performance of DDPMs. The proposed DDPM can be trained with higher resolution images, such as 128$\times$128 and 256$\times$256, compared to the original DDPM that generates 64$\times$64 pixels images. This work {involves} four steps: retinal vessel generation (1), retinal image synthesis (2), single image super-resolution (3), and retinal vessel segmentation (4). The proposed ReTree dataset was extensively evaluated quantitatively and qualitatively. In addition, it was compared with real retinal segmentation datasets. The results prove the superiority of our dataset over other manually collected datasets. In future works, we intend to improve the generation ability of DDPM, overcoming its limitations, mentioned in \ref{Limitations}. The proposed dataset and the source code of this work are available online for further evaluation and re-generating of the test results.

\ifpeerreview \else
\subsection*{Acknowledgments}
This work is partially supported by the Scientific and Technological Research Council of Turkey (TUBITAK) under the 2232 Outstanding Researchers program, Project No. 118C301. 

\subsection*{Availability of Source Code and Data}
Our developed dataset and source code is available at \url{https://github.com/AAleka/retree}.


\subsection*{Compliance with Ethical Standards}
This article does not contain any studies with human participants and/or animals performed by any of the authors

\subsection*{Conflict of Interest} 
The authors confirm that no actual or potential conflict of interest is related to this article.
\fi

\bibliographystyle{IEEEtran}
\small
\bibliography{references}

\begin{thebibliography}{10}
\providecommand{\url}[1]{#1}
\csname url@samestyle\endcsname
\providecommand{\newblock}{\relax}
\providecommand{\bibinfo}[2]{#2}
\providecommand{\BIBentrySTDinterwordspacing}{\spaceskip=0pt\relax}
\providecommand{\BIBentryALTinterwordstretchfactor}{4}
\providecommand{\BIBentryALTinterwordspacing}{\spaceskip=\fontdimen2\font plus
\BIBentryALTinterwordstretchfactor\fontdimen3\font minus
  \fontdimen4\font\relax}
\providecommand{\BIBforeignlanguage}[2]{{%
\expandafter\ifx\csname l@#1\endcsname\relax
\typeout{** WARNING: IEEEtran.bst: No hyphenation pattern has been}%
\typeout{** loaded for the language `#1'. Using the pattern for}%
\typeout{** the default language instead.}%
\else
\language=\csname l@#1\endcsname
\fi
#2}}
\providecommand{\BIBdecl}{\relax}
\BIBdecl

\bibitem{staal2004ridge}
J.~Staal, M.~D. Abr{\`a}moff, M.~Niemeijer, M.~A. Viergever, and
  B.~Van~Ginneken, ``Ridge-based vessel segmentation in color images of the
  retina,'' \emph{IEEE transactions on medical imaging}, vol.~23, no.~4, pp.
  501--509, 2004.

\bibitem{hoover2000locating}
A.~Hoover, V.~Kouznetsova, and M.~Goldbaum, ``Locating blood vessels in retinal
  images by piecewise threshold probing of a matched filter response,''
  \emph{IEEE Transactions on Medical imaging}, vol.~19, no.~3, pp. 203--210,
  2000.

\bibitem{fraz2012ensemble}
M.~M. Fraz, P.~Remagnino, A.~Hoppe, B.~Uyyanonvara, A.~R. Rudnicka, C.~G. Owen,
  and S.~A. Barman, ``An ensemble classification-based approach applied to
  retinal blood vessel segmentation,'' \emph{IEEE Transactions on Biomedical
  Engineering}, vol.~59, no.~9, pp. 2538--2548, 2012.

\bibitem{li2020iternet}
L.~Li, M.~Verma, Y.~Nakashima, H.~Nagahara, and R.~Kawasaki, ``Iternet: Retinal
  image segmentation utilizing structural redundancy in vessel networks,'' in
  \emph{Proceedings of the IEEE/CVF winter conference on applications of
  computer vision}, 2020, pp. 3656--3665.

\bibitem{ronneberger2015u}
O.~Ronneberger, P.~Fischer, and T.~Brox, ``U-net: Convolutional networks for
  biomedical image segmentation,'' in \emph{International Conference on Medical
  image computing and computer-assisted intervention}.\hskip 1em plus 0.5em
  minus 0.4em\relax Springer, 2015, pp. 234--241.

\bibitem{guo2022novel}
X.~Guo, X.~Lu, Q.~Lin, J.~Zhang, X.~Hu, and S.~Che, ``A novel retinal image
  generation model with the preservation of structural similarity and high
  resolution,'' \emph{Biomedical Signal Processing and Control}, vol.~78, p.
  104004, 2022.

\bibitem{andreini2021two}
P.~Andreini, G.~Ciano, S.~Bonechi, C.~Graziani, V.~Lachi, A.~Mecocci, A.~Sodi,
  F.~Scarselli, and M.~Bianchini, ``A two-stage gan for high-resolution retinal
  image generation and segmentation,'' \emph{Electronics}, vol.~11, no.~1,
  p.~60, 2021.

\bibitem{niu2021explainable}
Y.~Niu, L.~Gu, Y.~Zhao, and F.~Lu, ``Explainable diabetic retinopathy detection
  and retinal image generation,'' \emph{IEEE Journal of Biomedical and Health
  Informatics}, vol.~26, no.~1, pp. 44--55, 2021.

\bibitem{chen2021challenges}
H.~Chen, ``Challenges and corresponding solutions of generative adversarial
  networks (gans): a survey study,'' in \emph{Journal of Physics: Conference
  Series}, vol. 1827, no.~1.\hskip 1em plus 0.5em minus 0.4em\relax IOP
  Publishing, 2021, p. 012066.

\bibitem{sohl2015deep}
J.~Sohl-Dickstein, E.~Weiss, N.~Maheswaranathan, and S.~Ganguli, ``Deep
  unsupervised learning using nonequilibrium thermodynamics,'' in
  \emph{International Conference on Machine Learning}.\hskip 1em plus 0.5em
  minus 0.4em\relax PMLR, 2015, pp. 2256--2265.

\bibitem{heusel2017gans}
M.~Heusel, H.~Ramsauer, T.~Unterthiner, B.~Nessler, and S.~Hochreiter, ``Gans
  trained by a two time-scale update rule converge to a local nash
  equilibrium,'' \emph{Advances in neural information processing systems},
  vol.~30, 2017.

\bibitem{wang2004image}
Z.~Wang, A.~C. Bovik, H.~R. Sheikh, and E.~P. Simoncelli, ``Image quality
  assessment: from error visibility to structural similarity,'' \emph{IEEE
  transactions on image processing}, vol.~13, no.~4, pp. 600--612, 2004.

\bibitem{kim2022synthesizing}
M.~Kim, Y.~N. Kim, M.~Jang, J.~Hwang, H.-K. Kim, S.~C. Yoon, Y.~J. Kim, and
  N.~Kim, ``Synthesizing realistic high-resolution retina image by style-based
  generative adversarial network and its utilization,'' \emph{Scientific
  Reports}, vol.~12, no.~1, p. 17307, 2022.

\bibitem{liu2019synthesizing}
Y.-C. Liu, H.-H. Yang, C.-H. Huck~Yang, J.-H. Huang, M.~Tian, H.~Morikawa,
  Y.-C.~J. Tsai, and J.~Tegner, ``Synthesizing new retinal symptom images by
  multiple generative models,'' in \emph{Computer Vision--ACCV 2018 Workshops:
  14th Asian Conference on Computer Vision, Perth, Australia, December 2--6,
  2018, Revised Selected Papers 14}.\hskip 1em plus 0.5em minus 0.4em\relax
  Springer, 2019, pp. 235--250.

\bibitem{diaz2019retinal}
A.~Diaz-Pinto, A.~Colomer, V.~Naranjo, S.~Morales, Y.~Xu, and A.~F. Frangi,
  ``Retinal image synthesis and semi-supervised learning for glaucoma
  assessment,'' \emph{IEEE transactions on medical imaging}, vol.~38, no.~9,
  pp. 2211--2218, 2019.

\bibitem{yu2019retinal}
Z.~Yu, Q.~Xiang, J.~Meng, C.~Kou, Q.~Ren, and Y.~Lu, ``Retinal image synthesis
  from multiple-landmarks input with generative adversarial networks,''
  \emph{Biomedical engineering online}, vol.~18, no.~1, pp. 1--15, 2019.

\bibitem{appan2018retinal}
P.~Appan~K and J.~Sivaswamy, ``Retinal image synthesis for cad development,''
  in \emph{International Conference Image Analysis and Recognition}.\hskip 1em
  plus 0.5em minus 0.4em\relax Springer, 2018, pp. 613--621.

\bibitem{guibas2017synthetic}
J.~T. Guibas, T.~S. Virdi, and P.~S. Li, ``Synthetic medical images from dual
  generative adversarial networks,'' \emph{arXiv preprint arXiv:1709.01872},
  2017.

\bibitem{costa2017end}
P.~Costa, A.~Galdran, M.~I. Meyer, M.~Niemeijer, M.~Abr{\`a}moff, A.~M.
  Mendon{\c{c}}a, and A.~Campilho, ``End-to-end adversarial retinal image
  synthesis,'' \emph{IEEE transactions on medical imaging}, vol.~37, no.~3, pp.
  781--791, 2017.

\bibitem{costa2017towards}
P.~Costa, A.~Galdran, M.~I. Meyer, M.~D. Abramoff, M.~Niemeijer, A.~M.
  Mendon{\c{c}}a, and A.~Campilho, ``Towards adversarial retinal image
  synthesis,'' \emph{arXiv preprint arXiv:1701.08974}, 2017.

\bibitem{goodfellow2020generative}
I.~Goodfellow, J.~Pouget-Abadie, M.~Mirza, B.~Xu, D.~Warde-Farley, S.~Ozair,
  A.~Courville, and Y.~Bengio, ``Generative adversarial networks,''
  \emph{Communications of the ACM}, vol.~63, no.~11, pp. 139--144, 2020.

\bibitem{simonyan2014very}
K.~Simonyan and A.~Zisserman, ``Very deep convolutional networks for
  large-scale image recognition,'' \emph{arXiv preprint arXiv:1409.1556}, 2014.

\bibitem{ho2020denoising}
J.~Ho, A.~Jain, and P.~Abbeel, ``Denoising diffusion probabilistic models,''
  \emph{Advances in Neural Information Processing Systems}, vol.~33, pp.
  6840--6851, 2020.

\bibitem{song2020denoising}
J.~Song, C.~Meng, and S.~Ermon, ``Denoising diffusion implicit models,''
  \emph{arXiv preprint arXiv:2010.02502}, 2020.

\bibitem{choi2021ilvr}
J.~Choi, S.~Kim, Y.~Jeong, Y.~Gwon, and S.~Yoon, ``Ilvr: Conditioning method
  for denoising diffusion probabilistic models,'' \emph{arXiv preprint
  arXiv:2108.02938}, 2021.

\bibitem{nichol2021improved}
A.~Q. Nichol and P.~Dhariwal, ``Improved denoising diffusion probabilistic
  models,'' in \emph{International Conference on Machine Learning}.\hskip 1em
  plus 0.5em minus 0.4em\relax PMLR, 2021, pp. 8162--8171.

\bibitem{wang2018esrgan}
X.~Wang, K.~Yu, S.~Wu, J.~Gu, Y.~Liu, C.~Dong, Y.~Qiao, and C.~Change~Loy,
  ``Esrgan: Enhanced super-resolution generative adversarial networks,'' in
  \emph{Proceedings of the European conference on computer vision (ECCV)
  workshops}, 2018, pp. 0--0.

\bibitem{ledig2017photo}
C.~Ledig, L.~Theis, F.~Husz{\'a}r, J.~Caballero, A.~Cunningham, A.~Acosta,
  A.~Aitken, A.~Tejani, J.~Totz, Z.~Wang \emph{et~al.}, ``Photo-realistic
  single image super-resolution using a generative adversarial network,'' in
  \emph{Proceedings of the IEEE conference on computer vision and pattern
  recognition}, 2017, pp. 4681--4690.

\bibitem{alimanov2023hybrid}
A.~Alimanov, M.~B. Islam, and N.~F. Abubacker, ``A hybrid approach for retinal
  image super-resolution,'' \emph{Biomedical Engineering Advances}, p. 100099,
  2023.

\bibitem{qiu2022improved}
D.~Qiu, Y.~Cheng, and X.~Wang, ``Improved generative adversarial network for
  retinal image super-resolution,'' \emph{Computer Methods and Programs in
  Biomedicine}, p. 106995, 2022.

\bibitem{zhang2019attention}
S.~Zhang, H.~Fu, Y.~Yan, Y.~Zhang, Q.~Wu, M.~Yang, M.~Tan, and Y.~Xu,
  ``Attention guided network for retinal image segmentation,'' in
  \emph{International conference on medical image computing and
  computer-assisted intervention}.\hskip 1em plus 0.5em minus 0.4em\relax
  Springer, 2019, pp. 797--805.

\bibitem{fu2018joint}
H.~Fu, J.~Cheng, Y.~Xu, D.~W.~K. Wong, J.~Liu, and X.~Cao, ``Joint optic disc
  and cup segmentation based on multi-label deep network and polar
  transformation,'' \emph{IEEE transactions on medical imaging}, vol.~37,
  no.~7, pp. 1597--1605, 2018.

\bibitem{he2012guided}
K.~He, J.~Sun, and X.~Tang, ``Guided image filtering,'' \emph{IEEE transactions
  on pattern analysis and machine intelligence}, vol.~35, no.~6, pp.
  1397--1409, 2012.

\bibitem{jiang2021multi}
Y.~Jiang, W.~Liu, C.~Wu, and H.~Yao, ``Multi-scale and multi-branch
  convolutional neural network for retinal image segmentation,''
  \emph{Symmetry}, vol.~13, no.~3, p. 365, 2021.

\bibitem{li2021ta}
Y.~Li, J.~Yang, J.~Ni, A.~Elazab, and J.~Wu, ``Ta-net: Triple attention network
  for medical image segmentation,'' \emph{Computers in Biology and Medicine},
  vol. 137, p. 104836, 2021.

\bibitem{zhang2021pyramid}
J.~Zhang, Y.~Zhang, and X.~Xu, ``Pyramid u-net for retinal vessel
  segmentation,'' in \emph{ICASSP 2021-2021 IEEE International Conference on
  Acoustics, Speech and Signal Processing (ICASSP)}.\hskip 1em plus 0.5em minus
  0.4em\relax IEEE, 2021, pp. 1125--1129.

\bibitem{baranchuk2021label}
D.~Baranchuk, I.~Rubachev, A.~Voynov, V.~Khrulkov, and A.~Babenko,
  ``Label-efficient semantic segmentation with diffusion models,'' \emph{arXiv
  preprint arXiv:2112.03126}, 2021.

\bibitem{amit2021segdiff}
T.~Amit, E.~Nachmani, T.~Shaharbany, and L.~Wolf, ``Segdiff: Image segmentation
  with diffusion probabilistic models,'' \emph{arXiv preprint
  arXiv:2112.00390}, 2021.

\bibitem{alimanov2022retinal1}
A.~Alimanov and M.~B. Islam, ``Retinal image restoration and vessel
  segmentation using modified cycle-cbam and cbam-unet,'' in \emph{2022
  Innovations in Intelligent Systems and Applications Conference (ASYU)}.\hskip
  1em plus 0.5em minus 0.4em\relax IEEE, 2022, pp. 1--6.

\bibitem{hendrycks2016gaussian}
D.~Hendrycks and K.~Gimpel, ``Gaussian error linear units (gelus),''
  \emph{arXiv preprint arXiv:1606.08415}, 2016.

\bibitem{elfwing2018sigmoid}
S.~Elfwing, E.~Uchibe, and K.~Doya, ``Sigmoid-weighted linear units for neural
  network function approximation in reinforcement learning,'' \emph{Neural
  Networks}, vol. 107, pp. 3--11, 2018.

\bibitem{lee2021vision}
S.~H. Lee, S.~Lee, and B.~C. Song, ``Vision transformer for small-size
  datasets,'' \emph{arXiv preprint arXiv:2112.13492}, 2021.

\bibitem{fu2019evaluation}
H.~Fu, B.~Wang, J.~Shen, S.~Cui, Y.~Xu, J.~Liu, and L.~Shao, ``Evaluation of
  retinal image quality assessment networks in different color-spaces,'' in
  \emph{International Conference on Medical Image Computing and
  Computer-Assisted Intervention}.\hskip 1em plus 0.5em minus 0.4em\relax
  Springer, 2019, pp. 48--56.

\end{thebibliography}

\ifpeerreview \else



\begin{IEEEbiography}[{\includegraphics[width=1in,height=1.25in,clip,keepaspectratio]{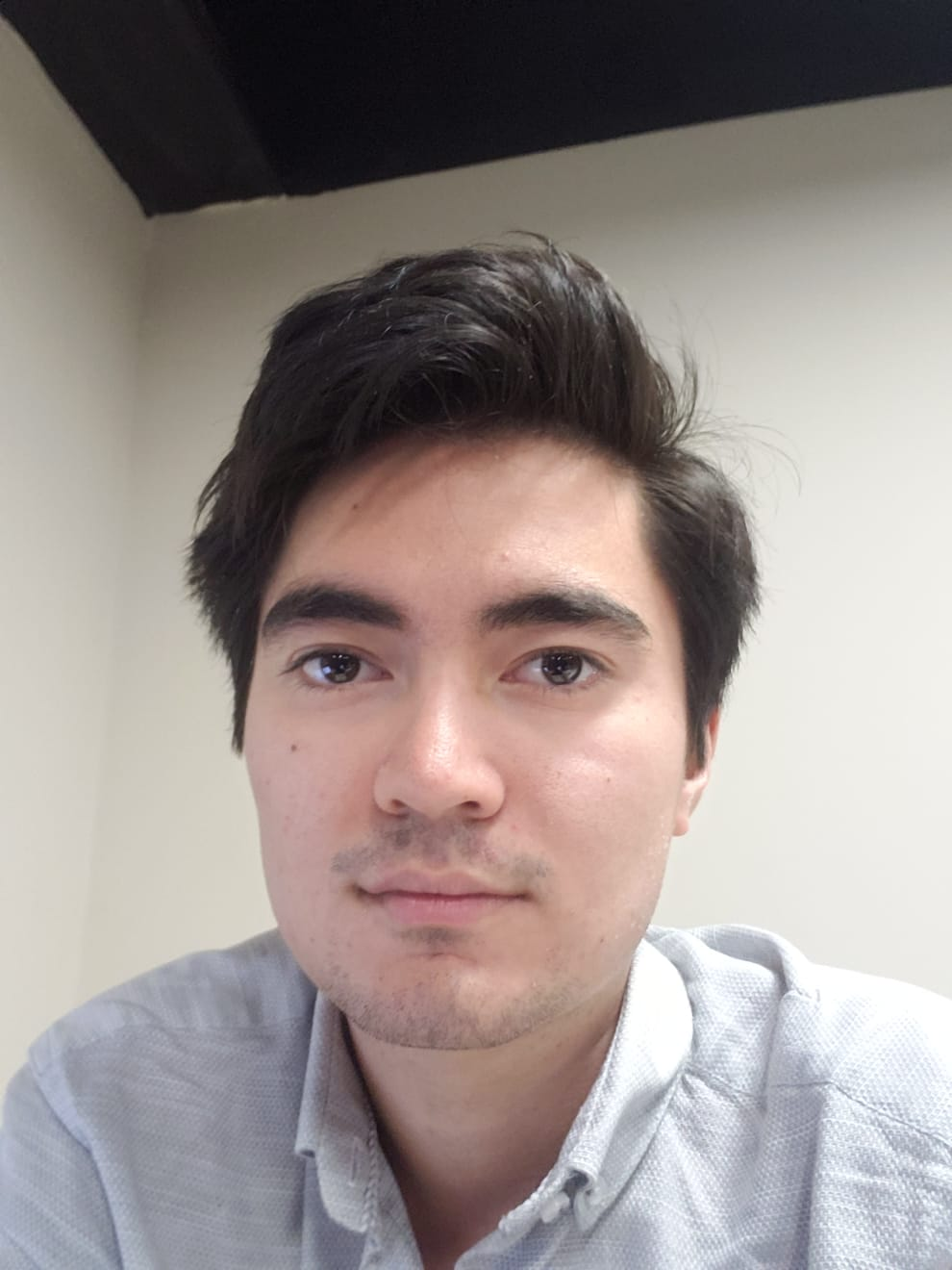}}]{Alnur Alimanov}
is an Artificial Intelligence Engineering master's student at Bahcesehir University in Istanbul, Turkey. He is a research assistant in a computer vision lab, funded by TUBITAK. He studied Mathematical Engineering for his bachelor's degree at International Information Technologies University in Almaty, Kazakhstan. He has several published papers related to retinal images, such as image enhancement, segmentation, and super-resolution. His current research interests are computer vision, natural language processing, deep learning, image and video processing, and medical image analysis.
\end{IEEEbiography}


\begin{IEEEbiography}[{\includegraphics[width=1in,height=1.25in,clip,keepaspectratio]{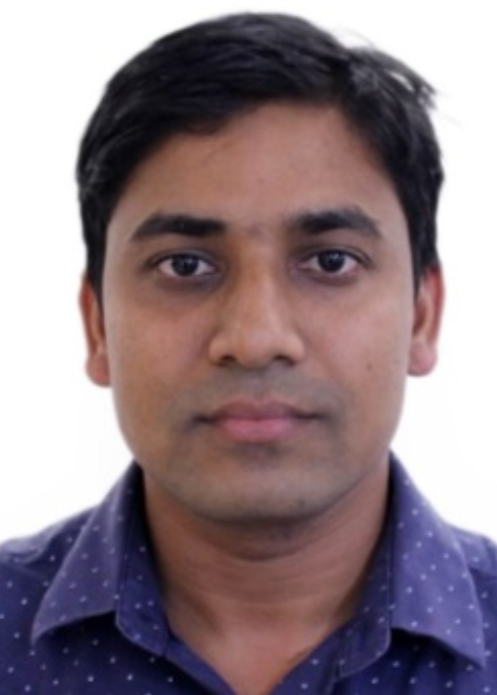}}]{Md Baharul Islam} is an Associate Professor of Computer Science at the College of Data Science \& Engineering of the American University of Malta and an Adjunct Professor of Computer Engineering at Bahcesehir University, Istanbul, Turkey. Prior he was a Postdoctoral Research Fellow at AI and Augmented Vision Lab at the University of Miami, USA. He completed his Ph.D. in Computer Science from Multimedia University in Malaysia, M.Sc. in Digital Media Technology from Nanyang Technological University in Singapore, and B.Sc. in Computer Science and Engineering from RUET, Bangladesh. Dr. Islam has more than 15 years of working experience in teaching and cutting-edge research in image processing and computer vision area. His current research interests lie in 3D stereoscopic media processing, computer vision, and AR/VR-based vision rehabilitation. Dr. Islam secured several (four) gold medals and six best research paper awards from international scientific competitions and conferences. His Ph.D. thesis was selected as the best Ph.D. thesis and received IEEE SPS Research Excellence Award 2018. He was also awarded the International Fellowship for Outstanding Young Researchers from TUBITAK in 2019. He authored/co-authored 80+ peer-reviewed research papers, including patents, journal articles, conference proceedings, and book chapters. Dr. Islam secured external research funds of about \$750,000.00. He leads an extensive research team, including postdocs and several Ph.D. and master’s students. He served as a program/technical committee member of many international conferences and workshops. Dr. Islam is an active IEEE Senior Member since 2018. 
\end{IEEEbiography}


\fi

\end{document}